\documentclass[12pt]{article}
\pdfoutput=1
\usepackage{wrapfig}
\usepackage{epsfig}
\usepackage[english]{babel}
\usepackage{hyperref}
\newcommand{\nc}{\newcommand}
\usepackage{latexsym}
\usepackage{amsmath}
\usepackage{amsfonts}
\usepackage{amssymb}
\begin{document}
\nc{\beq}{\begin{equation}} \nc{\eeq}{\end{equation}}
\nc{\beqa}{\begin{eqnarray}} \nc{\eeqa}{\end{eqnarray}}
\renewcommand{\thefootnote}{\fnsymbol{footnote}}

\vspace{0.5cm}

\begin{center}
{\bf \large LANDSCAPE VIEW AT THE EDGE OF\\[0.3cm] A MYSTERY\footnote[2]{Theory Vision talk at the Third annual conference on the LHC Physics, St.Petersburg, 5 September, 2015}}
\vspace{1cm}

{\bf \large  D.I.Kazakov}\vspace{0.5cm}

{\it Bogoliubov Laboratory of Theoretical Physics,\\
Joint Institute for Nuclear Research, Dubna\\[0.2cm]
Alikhanov Institute for Theoretical and Experimental Physics, Moscow\\[0.2cm]
Moscow Institute of Physics and Technology, Dolgoprudny}\vspace{0.6cm}

\abstract{The situation in particle physics after the discovery of the Higgs boson is discussed. Is the Standard Model consistent quantum field theory? Does it describe all experi\-men\-tal data? Are there any indications of physics beyond the SM?  Is there another scale except for the EW and the Planck ones? Is the SM of particles physics compatible with Cosmology?  New challenges of hadron physics: exotic hadrons and dense hadronic matter. Search for new physics, from the Higgs sector to  dark matter, supersymmetry, extra dimensions
and compositeness. What do we expect?  What are the main targets?
We try to answer the main questions and describe the key issues of possible new physics beyond the SM.
}
\end{center}


\section{Introduction: The Standard Theory} 
 With the launch of the LHC we approached the mystery land that lies beyond the TeV border line.  We do not know what is hidden there and our task is to be prepared and not to miss
 the new expected or unexpected phenomena. The guiding line here is our knowledge of physics at lower scales, first of all of the Standard Model of fundamental interactions which for our current understanding seems to be completed. Our search for possible new physics is based on the comparison of experimental data with the SM predictions.  From this point of view it is useful to look back at the SM  and remind its basic features.
 
 The Standard Model (Theory) is the gauge quantum field theory based on the following main principles:
 \begin{itemize}
 \item Three gauged symmetries $SU_c(3)\times SU_L(2)\times U_Y(1)$ corresponding to the strong, weak and electromagnetic interactions, respectively;
 \item Three families of quarks and leptons  belonging to the representations  ($3\times 2, 3\times 1, 1\times 2, 1\times 1$) of the gauge groups;
  \item The Brout-Englert-Higgs mechanism of spontaneous EW symmetry breaking  accom\-panied with the Higgs boson;
  \item The CKM and PMNS mixing matrices of flavours;
  \item The CP violation via the phase factors in these matrices;
  \item Confinement of quarks and gluons inside hadrons;
  \item The Baryon and Lepton number conservation;
 \item  The CPT invariance leading to the existence of antimatter.
\end{itemize} 
 The ST principles allow:
 \begin{itemize}
\item[-]Extra families of quarks and leptons  ---  seems to be excluded experimentally already;
\item[-]Presence or absence of right-handed neutrino --- still unclear;
\item[-]Majorana/Dirac nature of neutrino --- the Majorana mass is slightly beyond the SM;
\item[-]Extra Higgs bosons --- Already beyond but in the spirit of the SM.
 \end{itemize}
 The main questions to the Standard Theory (ST) can be formulated as:
 \begin{itemize} 
\item[$\blacktriangleright$] Is it self consistent?
\item[$\blacktriangleright$] Does it describe all experimental data?
\item[$\blacktriangleright$] Are there any indications of physics beyond the SM?
\item[$\blacktriangleright$] Is there another scale except for the EW and the Planck ones?
\item[$\blacktriangleright$] Is it compatible with Cosmology?
 \end{itemize}
 There are also many "why's" and "how's":\vspace{0.5cm}
 
 \begin{tabular}{l|l} \vspace{0.3cm}
 \hspace*{2cm}Why's & \hspace{2cm}How's\\ 
 why  $SU(3)\times SU(2) \times U(1)$ ? & how does confinement actually work?\\
 why 3 generations? &  how does the quark-hadron phase transition happen?\\
 why quark-lepton symmetry? & how do neutrinos get a mass?\\
 why V-A weak interaction? & how does CP violation occur in the Universe? \\
 why L-R asymmetry? &  how to protect the SM from would be heavy\\
 why B \& L conservation? &  scale  physics?\\ etc &
  \end{tabular}
  
In what follows we will try to answer the main questions and describe the key issues of possible physics beyond the SM.

\section{Is the SM consistent quantum field theory?}
\subsection{Ghosts}
\begin{figure}[ht]\vspace{0.5cm}
\leavevmode\hspace{1cm}
\includegraphics[width=0.40\textwidth]{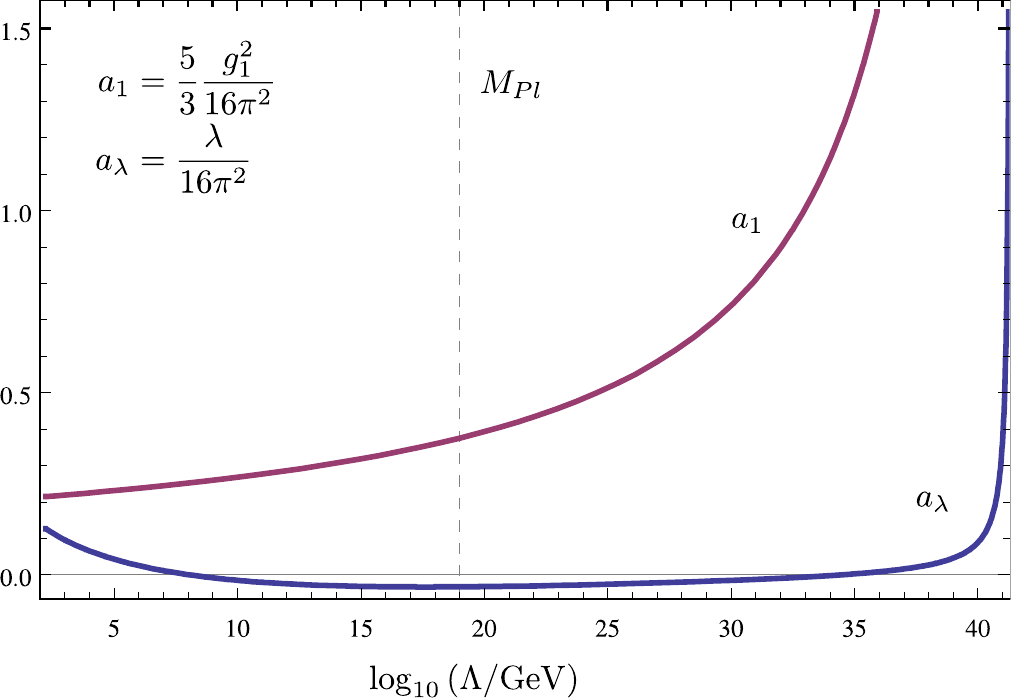}\vspace{-4.2cm}

 \hspace{9.3cm}
 \includegraphics[width=0.40\textwidth]{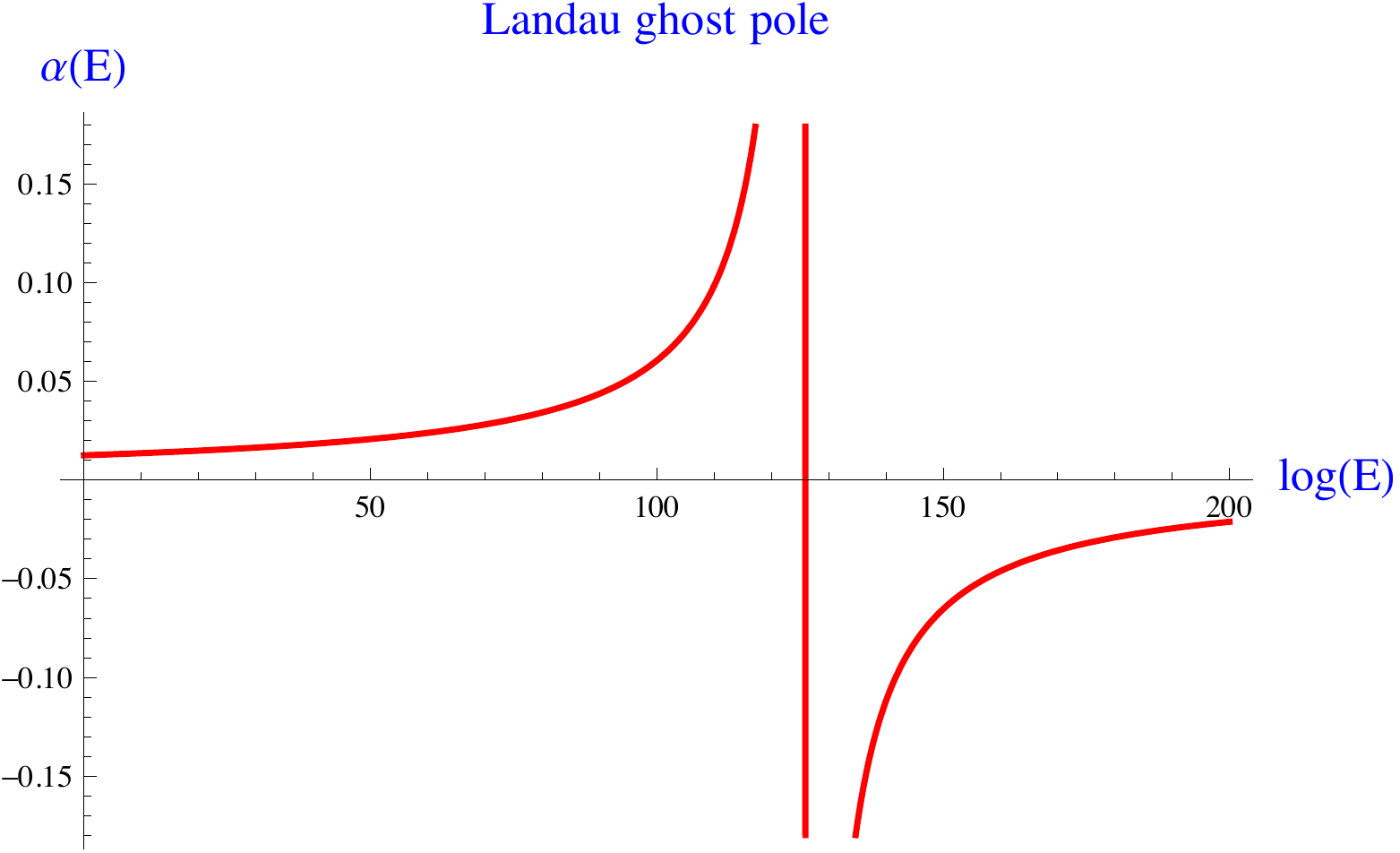}\vspace{1cm}
\caption{The running of the U(1) and the Higgs couplings (left) and the Landau ghost pole (right)}
\label{pole}
\end{figure}

For a long time the known property of the SM is that the running couplings possess the Landau ghost poles at high energies. This is true  for the U(1) coupling and for the  Higgs coupling,  but \dots beyond the Planck scale, as shown in Fig.\ref{pole}.
The Landau pole has a wrong sign residue that indicates the presence of unphysical ghost fields - intrinsic problem and inconsistency of a theory~\cite{Landau}. The one loop expression for the hyper charge coupling in the SM 
\beq
\alpha_1(Q^2)=\frac{\alpha_{10}}{1-\frac{41}{10}\frac{\alpha_{10}}{4\pi}\log(Q^2/M_Z^2)}
\eeq
possesses the ghost pole at  the scale $Q^*=M_Ze^{\frac{20\pi}{41\alpha_{10}}}\sim 10^{41}\ GeV$. It is far beyond the Planck scale and one may ignore it assuming that the Planck scale quantum gravity will change the situation. However, quantum gravity is still lacking and the presence of ghosts is intrinsically dangerous independently of the scale where they appear.
The situation may change in GUTs due to new heavy fields at the GUT scale. In any case, this  requires modification of the ST at VERY high energies.

\subsection{Anomalies}
\begin{figure}[ht]
\begin{center}
\leavevmode
\includegraphics[width=0.25\textwidth]{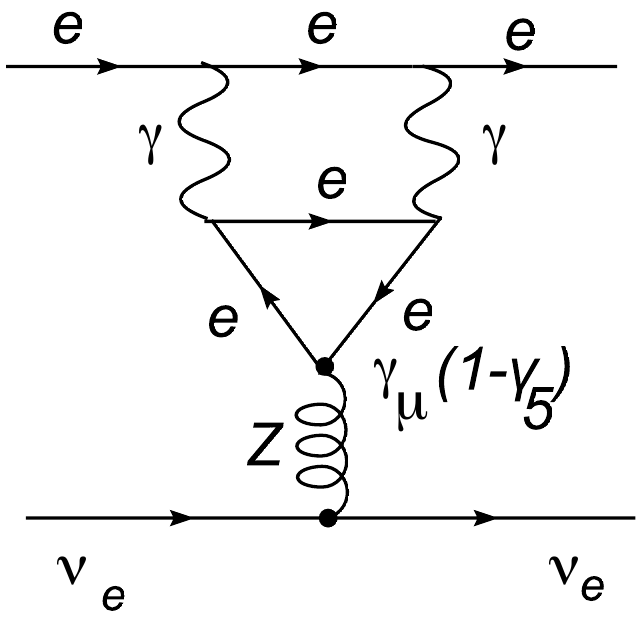}
\end{center}
\caption{The chiral anomaly diagram in the electron-neutrino scattering amplitude}
\label{anomaly}
\end{figure}

\begin{figure}[ht]
\begin{center}
\leavevmode
\includegraphics[width=0.75\textwidth]{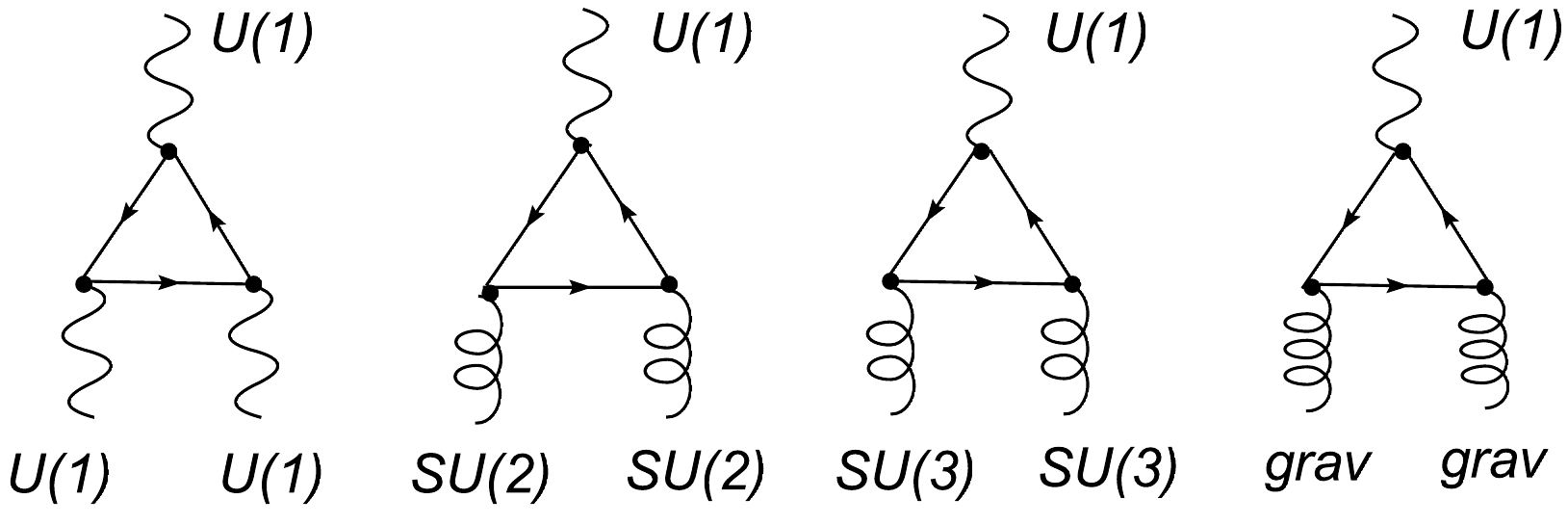}
\end{center}
\caption{The gauge anomalies in the SM}
\label{gaugeanomaly}
\end{figure}

As is well known, in quantum theories there may exist anomalies that can ruin the theory. In the SM there 
is a set of  quantum anomalies. A famous example is the triangle chiral anomaly~\cite{Anomaly}. Its contribution to the electron-neutrino scattering amplitude is shown in Fig.\ref{anomaly}. It would destroy renormalizability if not cancelled among quarks and leptons. The other anomalies existing in the SM are shown in Fig.\ref{gaugeanomaly}~\cite{AnomSM}. 
Fortunately, they are all canceled in the SM for each generation of quarks and leptons, as can be seen from expressions below.
$$\begin{array}{l}TrY^3=3\ (\frac{1}{27}+\frac{1}{27}-\frac{64}{27}+\frac{8}{27})-1-1+8=0,\\ \hspace*{1.5cm}\uparrow\\
\hspace*{0.8cm} color\ \ \  u_L \ \ \ d_L \ \ \ \  u_R \ \ d_R \hspace{0.5cm} \nu_L\ \ \ e_L \ \ e_R
\end{array}$$
$$Tr Y_L=3(\frac 13+\frac 13)-1-1=0,$$
$$Tr Y_q=3(\frac 13+\frac 13-\frac 43-(-\frac 23))=0,$$
$$TrY=3(\frac 13+\frac 13-\frac 43-(-\frac 23))-1-1-(-2)=0.$$
Thus, the cancellation of anomalies requires the quark-lepton symmetry. Probably, 
this is a hint towards the Grand Unified Theories.

\subsection{Vacuum stability}
\begin{figure}[ht]
\begin{center}
\leavevmode
\includegraphics[width=0.36\textwidth]{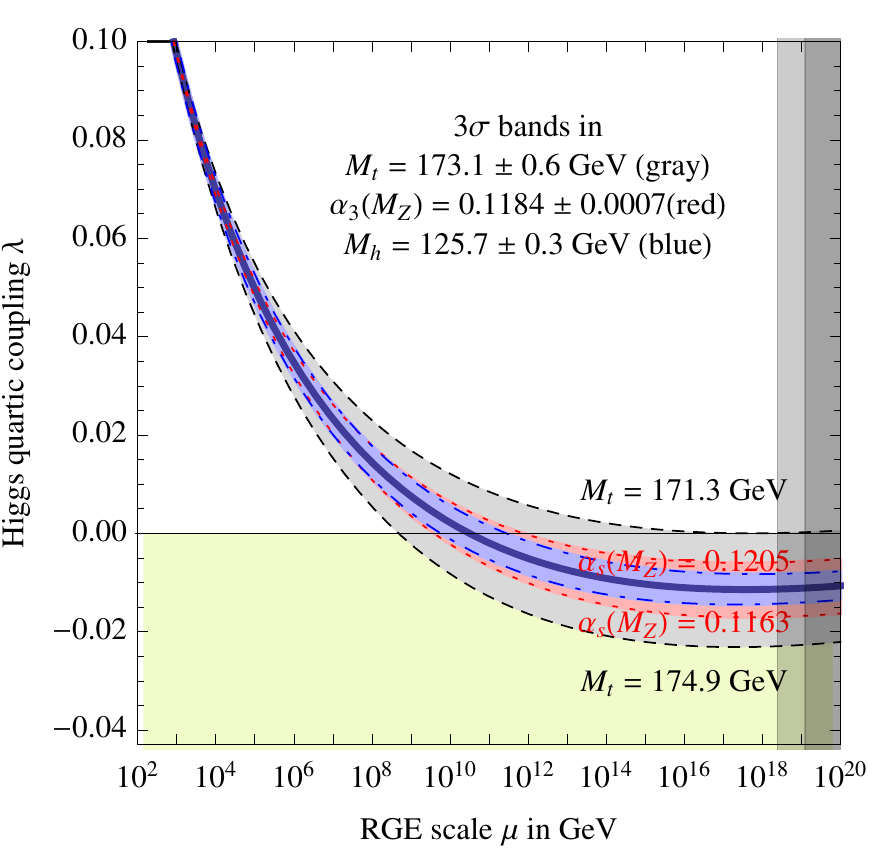}\hspace{1cm}
\includegraphics[width=0.35\textwidth]{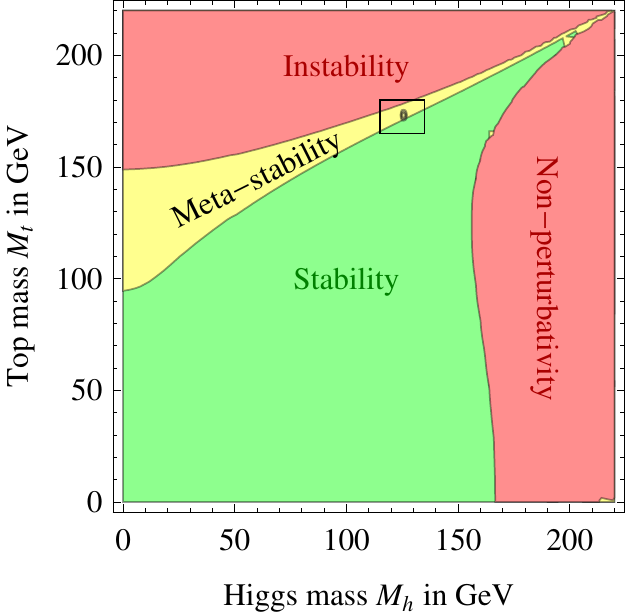}
\end{center}
\caption{The running of the Higgs coupling (left) and the stability of the EW vacuum as a function of
the Higgs and top masses}
\label{higgscoupling}
\end{figure}

\begin{figure}[ht]
\begin{center}
\leavevmode
\includegraphics[width=0.35\textwidth]{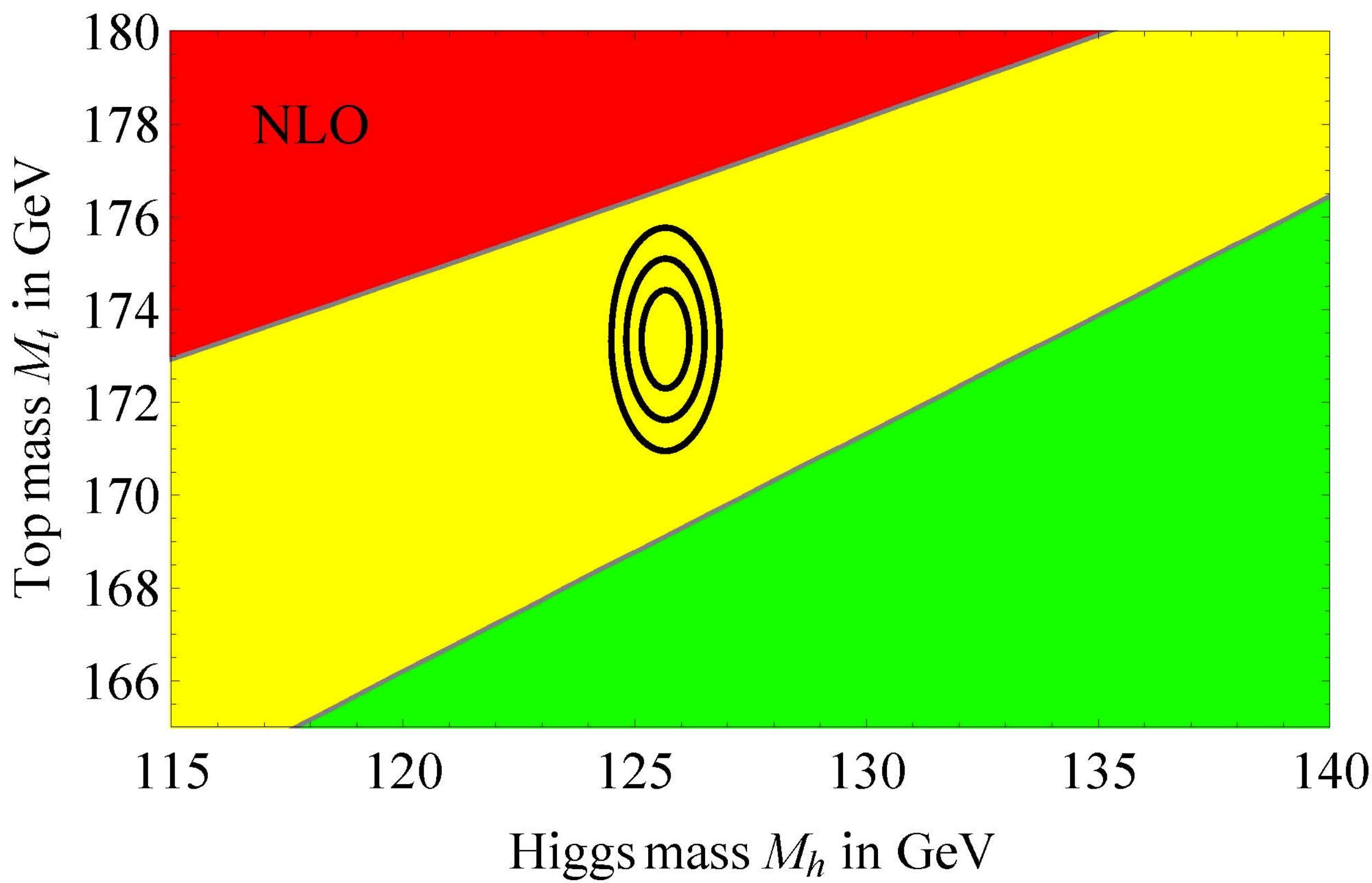}\hspace{1cm}
\includegraphics[width=0.33\textwidth]{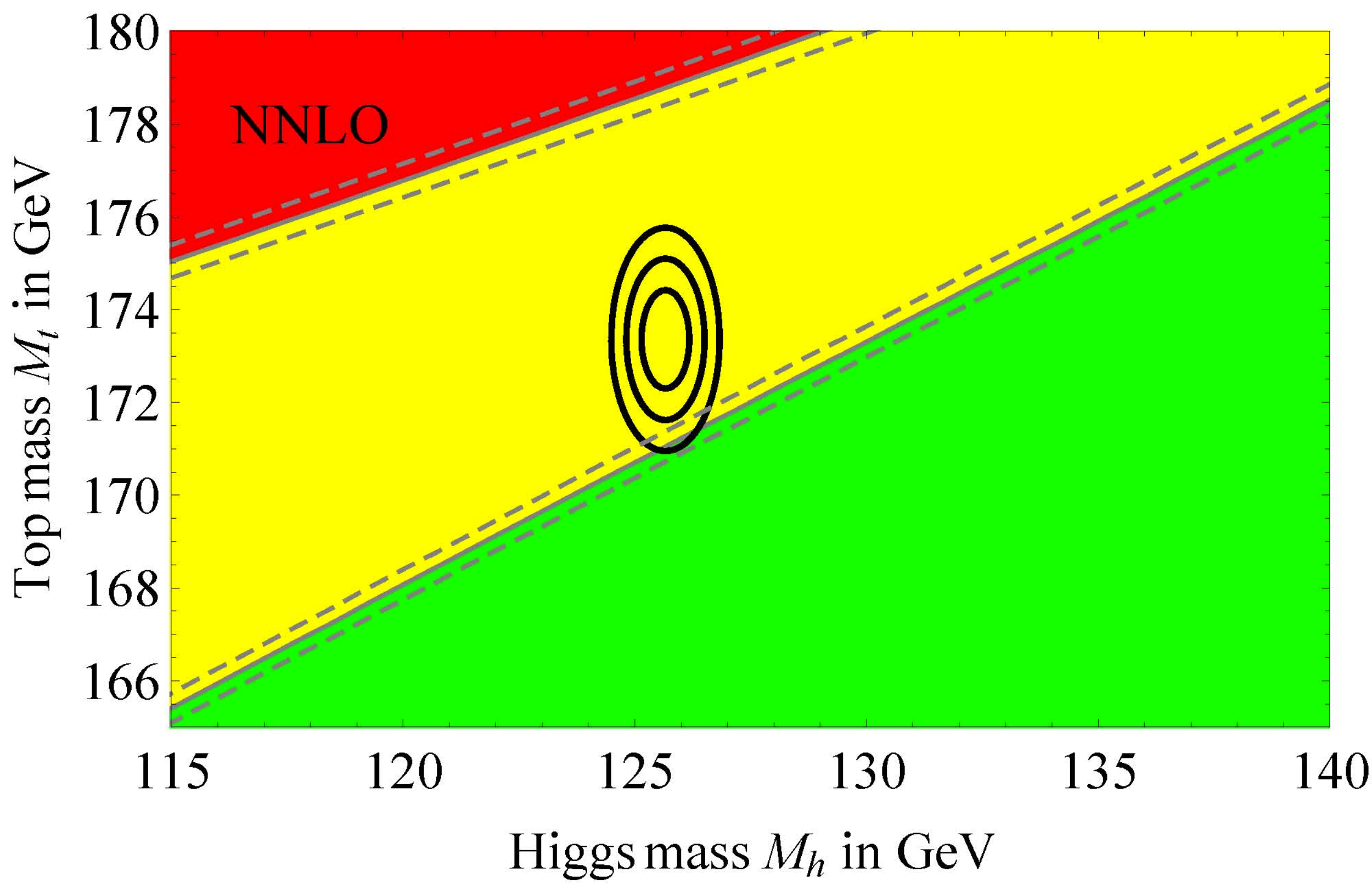}
\end{center}
\caption{The vacuum stability point at the NLO and NNLO}
\label{nnlo}
\end{figure}

Quantum corrections can make the vacuum unstable. Moreover, the whole construction of the SM may be in trouble being metastable or even unstable. This is related to the Higgs potential which at the tree level contains quadratic and quartic terms. The quartic coupling due to the radiative corrections depends on a scale and at some scale might change the sign, thus making the EW vacuum unstable. Indeed, it may happen at high energy scale, as shown in Fig.\ref{higgscoupling} (left)\cite{vacuum}. The situation crucially depends on the top and Higgs mass values and requires severe fine-tuning and accuracy (see Fig.\ref{higgscoupling} (right)).  It seems that we are sitting just at the border line with the top quark and the Higgs boson 
masses specially adjusted. However, the account of the next-to-leading order corrections is essential and shifts the position towards the stability region, as can be seen in Fig.\ref{nnlo}\cite{highorder}.

In the case when the EW vacuum is indeed metastable the right question to ask would be: what is the life-time of the ground state? If it is bigger than the life-time of the Universe, it is still fine for the SM. Still the situation requires some caution. The way out might be the  new physics  at higher scale. One example is supersymmetry: in this case the scalar potential is $V_{SUSY}=|F|^2+|D|^2\geq 0$~\cite{WB}. If SUSY is broken the potential can get negative corrections though the quartic scalar coupling remains to be positive. The  second example is the extended Higgs sector.
Several Higgs fields with several Higgs-like couplings  push the smallest coupling up 
(might have also several minima). The third example is provided by  GUTs. In a unified theory the Higgs coupling may be attracted by the gauge coupling, thus stabilizing the potential. Note that in all these cases one has an extension of the SM at high energies.

\subsection{Scale stability}
\begin{figure}[ht]
\begin{center}
\leavevmode
\includegraphics[width=0.5\textwidth]{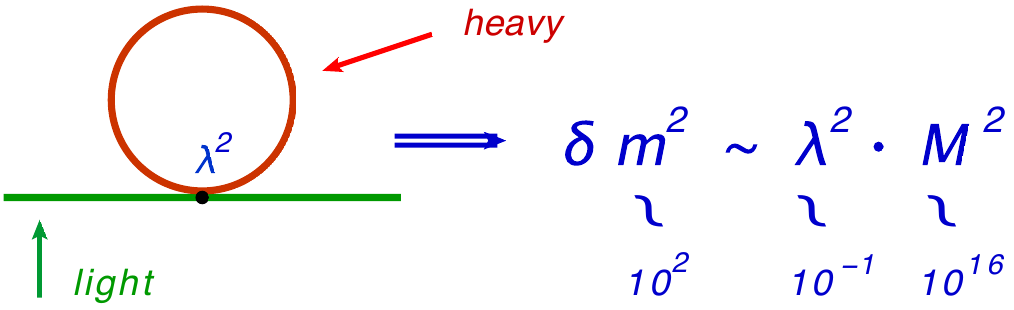}
\end{center}
\caption{Radiative correction to the Higgs mass due to heavy particles }
\label{corr}
\end{figure}

New physics at high energy scale  may destroy the EW scale of the SM. Indeed, the masses of quarks and leptons and the masses of gauge bosons in the SM are protected versus the radiative corrections originating from heavy new physics due to the gauge invariance. However, this is not true for the Higgs mass.  The Higgs sector is not protected by any symmetry. Quantum corrections to the Higgs potential from to New physics (see Fig.\ref{corr}) are proportional to the heavy mass squared.  These huge corrections would destroy the light Higgs potential  and eventually the light EW scale of the SM. This creates a hierarchy problem: the coexistence of the light and heavy scales $\frac{m_H}{m_{GUT}}\sim 10^{-14}$ and requires modification of the SM.
 \begin{figure}[ht]
\begin{center}
\leavevmode
\includegraphics[width=0.5\textwidth]{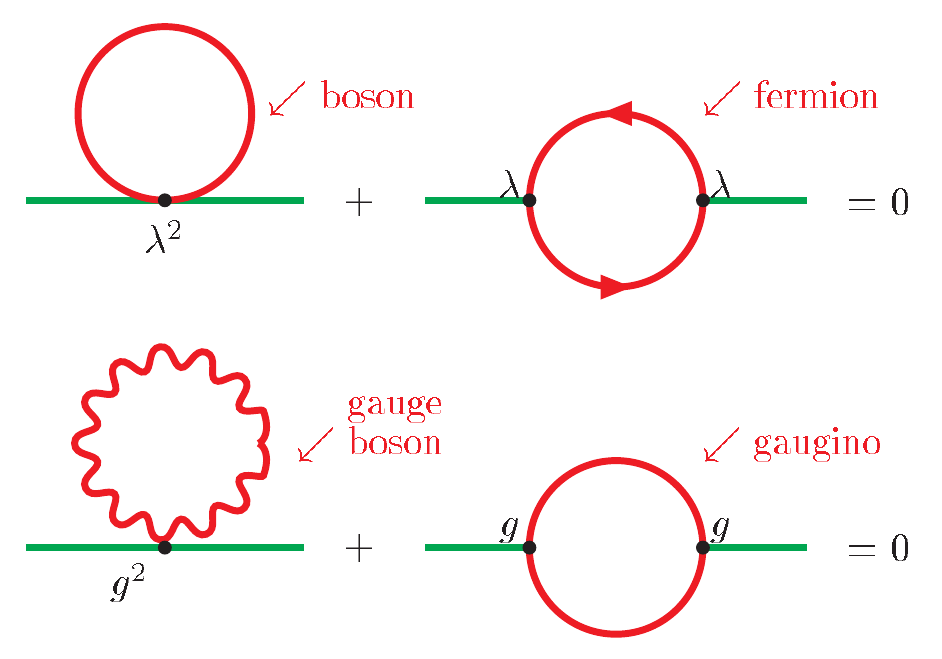}
\end{center}
\caption{Cancellation of the radiative correction to the Higgs mass due to super partners}
\label{cancel}
\end{figure} 

The way out again might be the  new physics  at higher scale. Two suggestions are popular.
The first one is supersymmetry at TeV scale. In this case, the unwanted radiative corrections are canceled by super partners of the corresponding particles, as it is shown in Fig.\ref{cancel}~\cite{Kazakov}. This cancellation is true up to the SUSY breaking scale. If $m_{SUSY}\sim 1$ TeV, the light Higgs mass is protected, which suggests an approximate scale of low energy supersymmetry. If, on the contrary,  $m_{SUSY}\geq 1$ TeV, one has the so-called little hierarchy problem that requires the fine tuning of the SUSY parameters~\cite{Littlehierarchy}.

\begin{figure}[ht]
\begin{center}
\leavevmode
\includegraphics[width=0.5\textwidth]{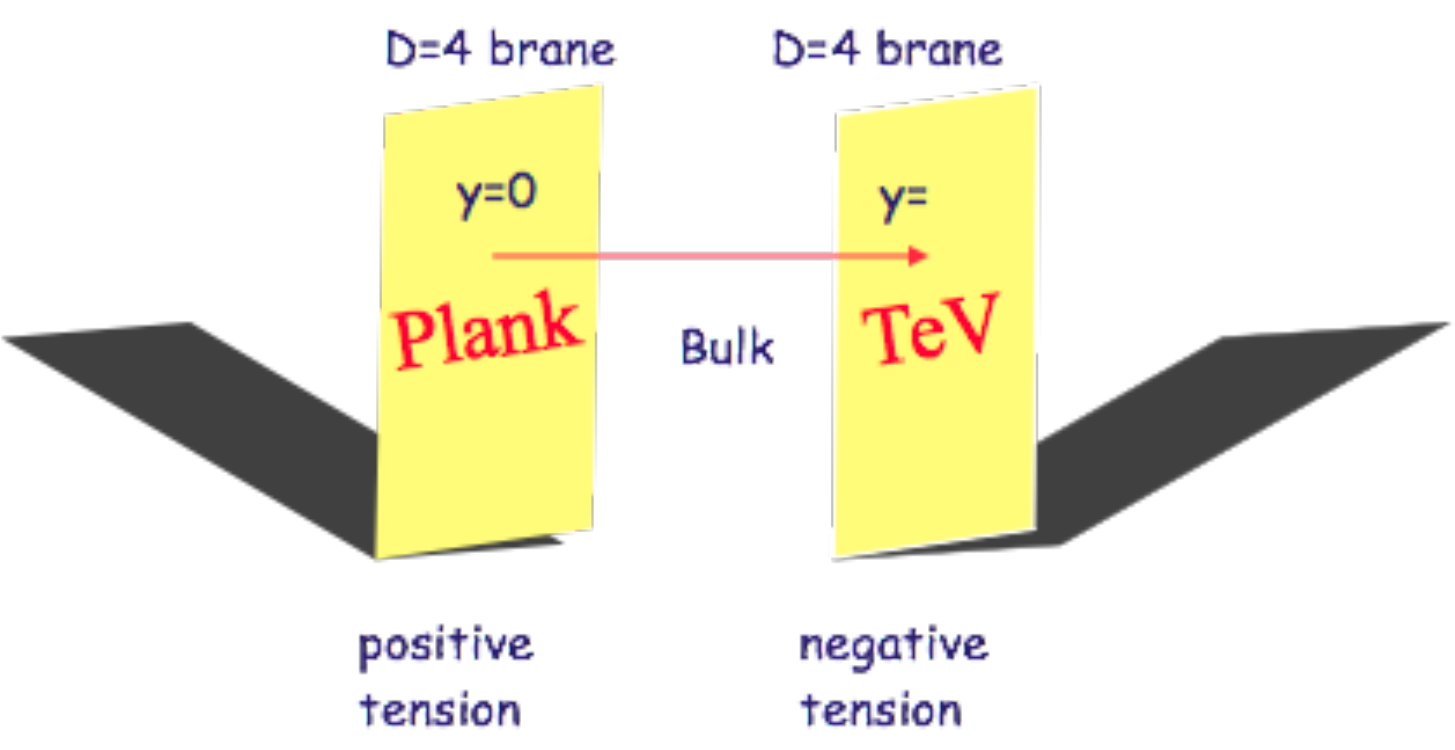}
\end{center}
\caption{The Randall-Sundrum type brane world construction}
\label{brane}
\end{figure}

The other proposal to solve the hierarchy problem is related to the extra dimensional theories. In this case
the hierarchy is achieved due to the wrap factor which appears while going from the so-called Planck brane
to the physical brane (Fig.\ref{brane}).  
 In the Randall-Sundrum brane world construction~\cite{RS} the gravity scale at the Planck brane and the TeV brane are related by
\beq M^2_{Pl}=\frac{M^3}{k}(e^{2k\pi R}-1).\eeq
As a result the gravity scale at the TeV brane, $M$, might be small enough not to create the hierarchy problem.
Whether any of these scenarios are realized in Nature is unclear.

\section{Does the ST describe all experimental data?}

\begin{figure}[ht]
\begin{center}
\leavevmode
\includegraphics[width=0.45\textwidth]{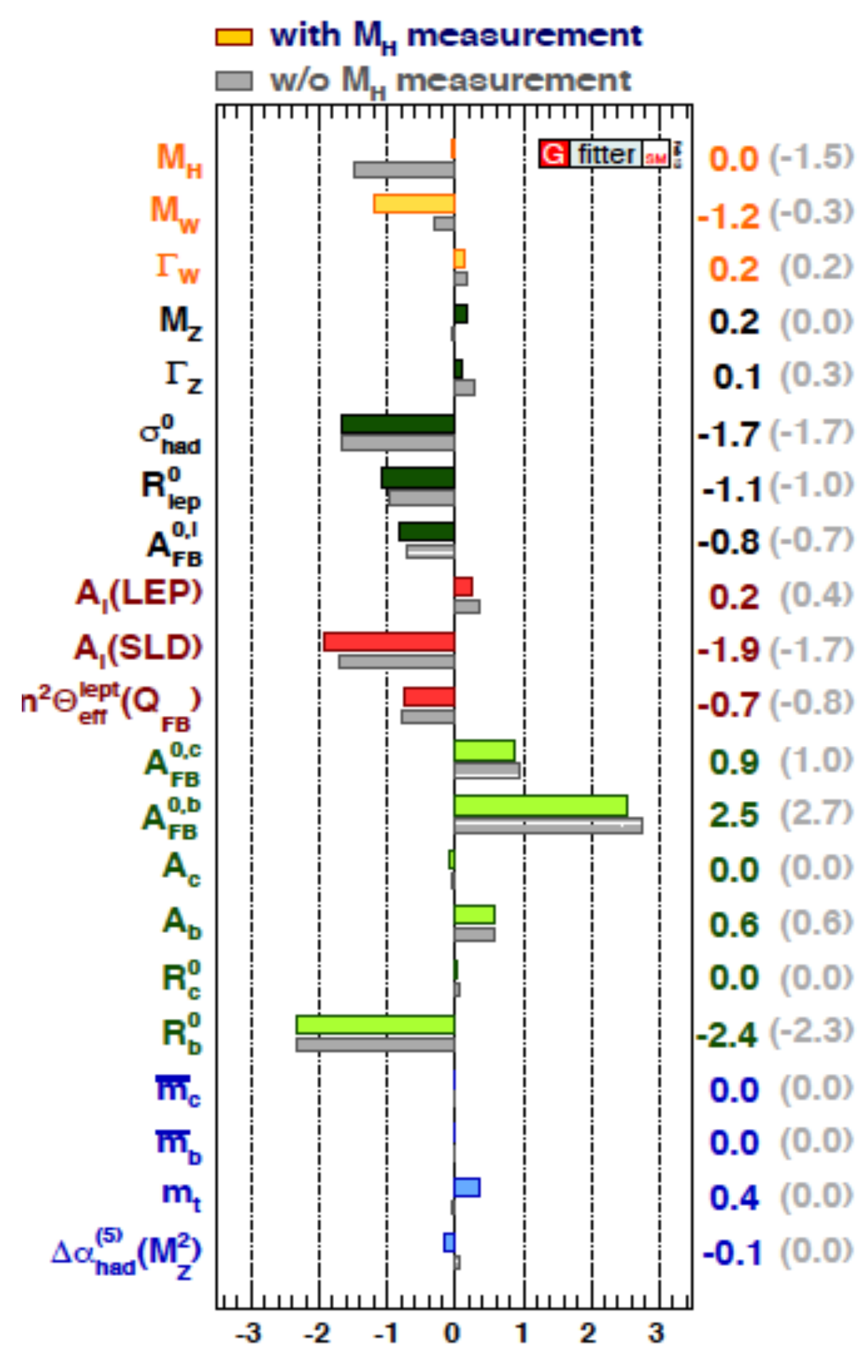}
\includegraphics[width=0.45\textwidth]{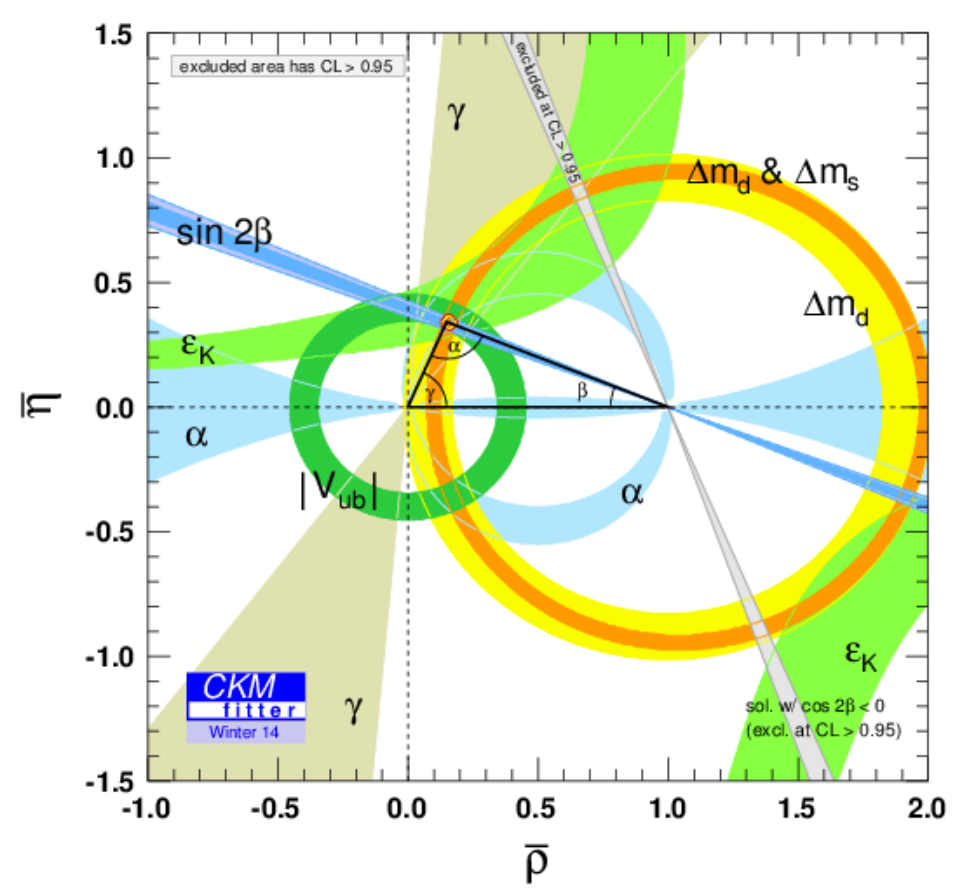}
\end{center}
\caption{The pool of the EW data (left) and the flavour observables (right)}
\label{EWpool}
\end{figure}

Remarkable success of the SM in describing practically all experimental data in particle physics manifests itself in a pool of EW observables (see Fig.\ref{EWpool} left)~\cite{EWPool}. Almost everywhere one has agreement   within 1-2 standard deviations. The only exception is the forward-backward asymmetries in LEP data, the long ignored problem usually attributed to the analysis. The same is true for the flavor observables  (see Fig.\ref{EWpool} right)~\cite{Flavor}.
One has to admit very impressive progress achieved in the last decade in the  EW and QCD perturbative calculations (see e.g. \cite{PTCalc}). This became possible due to the development of new techniques and computer codes for multi-loop and multi-leg calculations. Today the accuracy of theoretical calculations competes with that of experimental data and further progress is on the way in both the cases.

\begin{figure}[ht]
\begin{center}
\leavevmode
\includegraphics[width=0.40\textwidth]{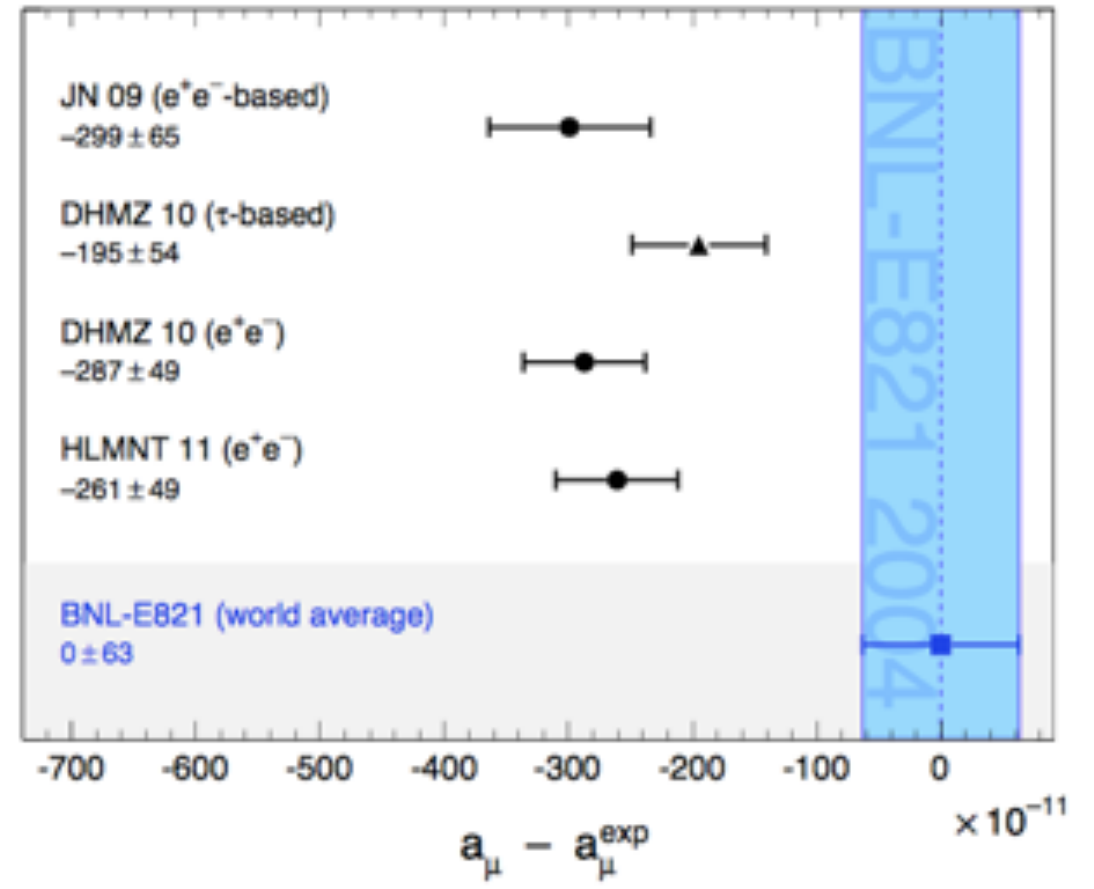}
\end{center}\vspace{-0.2cm}
\caption{The difference between the experimental and theoretical values of $a_\mu$}
\label{g-2}
\end{figure}

For years the pain in the neck remains an almost 3 $\sigma$ discrepancy in the anomalous magnetic moment of muon, the  $a_\mu={(g-2)}/2$, as illustrated in Fig.\ref{g-2}~\cite{g2}. The attempts to fill the gap with the new physics contributions are not very successful due to the heaviness of the experimentally allowed new physics. The reason is that the new physics contributions come from the virtual particles in the loop and 
these diagrams are suppressed by the inverse mass squared of the intermediate particles. Though this explanation is still possible, the main hopes in resolving the $a_\mu$ puzzle are related with the still inaccurate strong interaction contribution or with the new experiment which is on the way.

The other discrepancy  which attracted  attention is the value of the CKM mixing matrix element
$V_{ub}$  measured recently\cite{Vub}. This quantity is slightly different when measured in inclusive and exclusive processes. The resolution of this puzzle presumably lies in the theoretical interpretation.

Looking at the other observables and unsolved problems one has to mention the strong CP problem which despite being known for several decades still has not found its solution. The elegant way of resolving it with the help of the axion field still lacks experimental confirmation. The existing models with the axion field allow
almost invisible light particle leaving small chances for its detection~\cite{axion}. Another field where the new physics might appear is the rare decays. Here, despite some hopes connected in particular with the $B_s\to \mu\mu$ decay, everything looks fine for the SM so far~\cite{rare}. QCD is another huge area of activity. It also looks fine, although the spin crisis related to the spin of the proton is still unresolved. Presumably, it is related to parton distributions. Relatively new activity with the generalized parton distributions depending on momentum transfer opens a new field for the check of the SM~\cite{PartonD}. At last, the neutrino physics attracts much attention in recent and coming years. It seems that the neutrino masses and mixings look fine so far but still  need to be clarified. The nature of neutrino (Dirac or Majorana) remains the major puzzle in this field.

\begin{figure}[ht]
\begin{center}
\leavevmode
\includegraphics[width=0.35\textwidth]{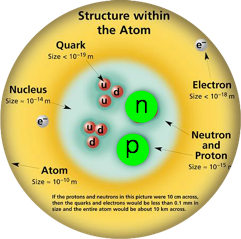}
\includegraphics[width=0.07\textwidth]{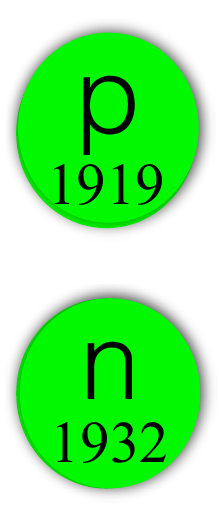}\hspace{0.6cm}
\includegraphics[width=0.50\textwidth]{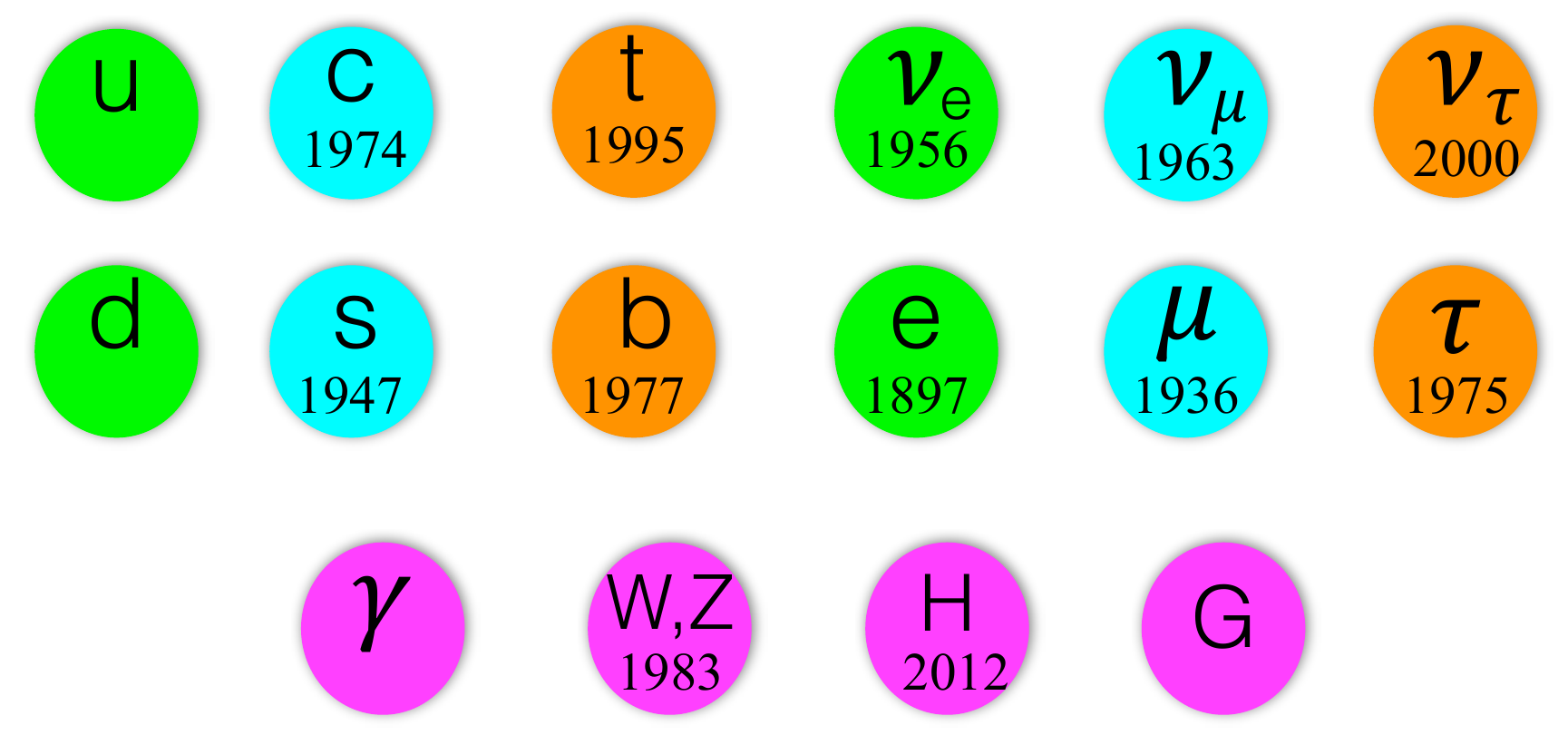}
\end{center}
\caption{The structure of the atom (left) and the families of quarks and leptons and the force carriers (right)}
\label{atom}
\end{figure}

In this situation one may wonder why do we talk about new physics at all. Everything seems to be described by the SM. It is useful to look back in history and find an analogy of the modern situation. For this purpose, let us go back to the middle of the 20th century. This was the world of a single generation of particles. Indeed, the world around us is made of the first generation. In the middle of the  20th century we had the following set of elementary particles: proton, neutron, electron and later neutrino. The structure of an atom was described in detail (see Fig.\ref{atom} left). Who expected new physics to come? And at which scale?

We know what happened next: the muon was discovered in cosmic rays.  It was first considered as a heavy electron and later was recognized as the beginning of the 2nd generation. Then the $K$-meson
appeared, the strange particle. The following up discoveries of new hadrons at accelerators triggered the
invention of the quark model and everything looked OK again. Then some problems with suppression of the flavour changing  transitions appeared which were resolved with the help of the GIM mechanism~\cite{GIM}. The subsequent discovery of J/Psi completed the 2-nd generation with the charm quark. This second generation looked artificial and unexplained; however, something was missing since the CP-violation was discovered and called for the interpretation. The Kobayashi-Maskawa  mixing matrix  gave a hint for the 3rd
generation and here we are. Discoveries of the force carriers and eventually of the Higgs boson already in the 21st century completed our picture, as shown in Fig.\ref{atom} (right). Only the gravitational force with the graviton as a carrier still stands aside. Let me repeat, who expected this in the middle of the 20th century?

There were, however, unanswered questions. The challenge came from
astrophysics and cosmology. Being at the  earlier stage of the development they puzzled particle physics with the major problems of  the baryon asymmetry of the Universe and the
description of the Dark Matter known already at that time. Remarkably that these problems are still not resolved within the SM today. 

\section{Is there another scale except for the EW and Planck ones?}

\begin{figure}[ht]
\begin{center}
\leavevmode
\includegraphics[width=0.95\textwidth]{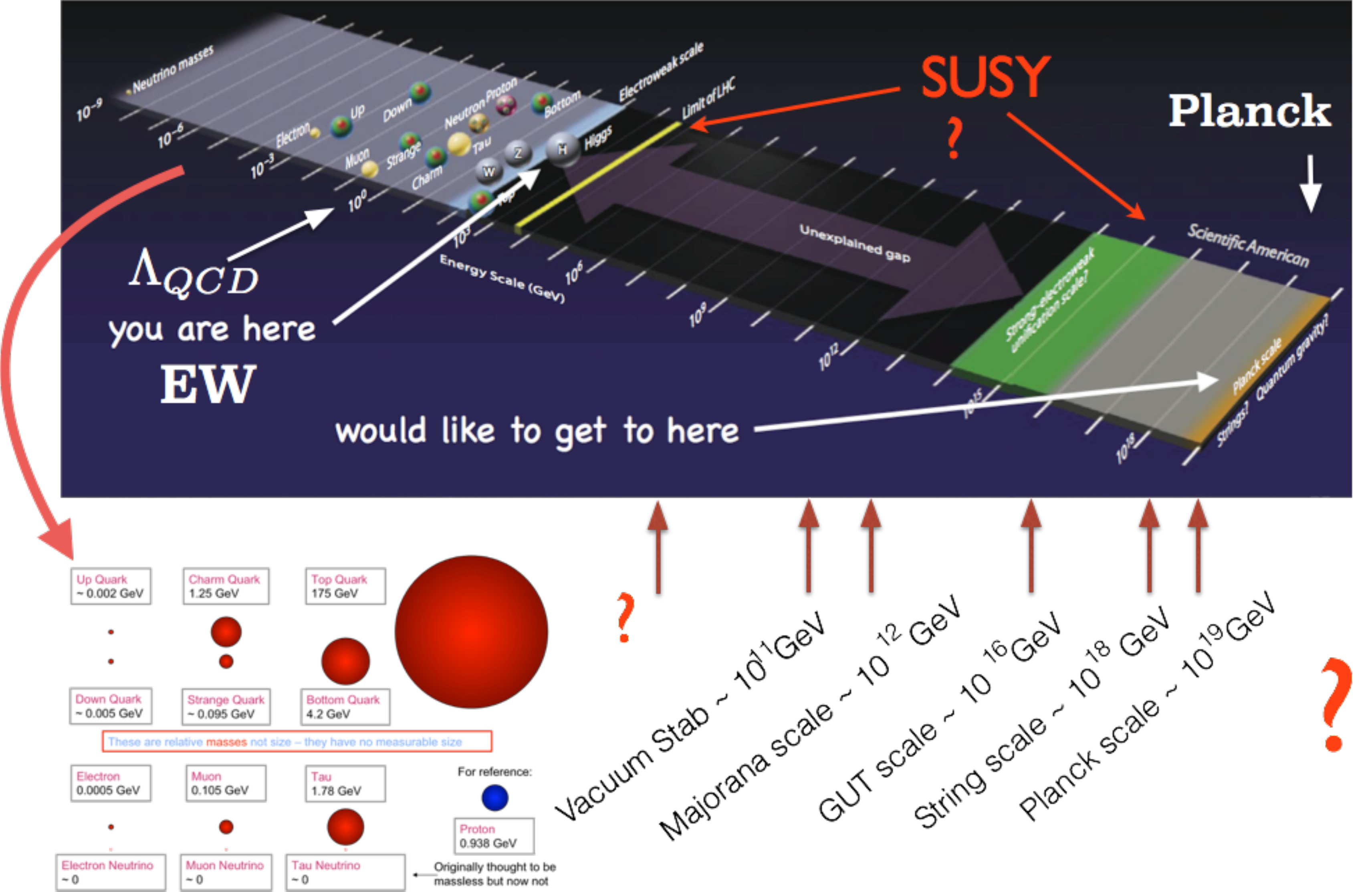}
\end{center}
\caption{The high energy physics panorama and possible scales of the new physics}
\label{scale}
\end{figure}

The expectations of the new physics inevitably lead to the question of the scale. Is there any new scale between the EW and the Planck ones? Many new phenomena assume the existence of the proper scales.
The forseeble panorama of high energy physics is shown in Fig.\ref{scale}~\cite{scale}. 
First of all there is the EW scale. All the masses (except for the top quark and the Higgs boson) lie below this scale and form a random pattern as of today. Then there is $\Lambda_{QCD}$ which is not a fundamental scale but plays an essential role in strong interactions. Moving down from the Planck scale we have subsequently the hypothetical string scale, the GUT scale, the Majorana scale, the vacuum stability scale, probably some others like the Pechei-Queen scale, etc. Somewhere in between is  the foreseen SUSY scale. There might also be  the scale of  extra dimensions positioned anywhere.
 
 What is true of this picture? Is there anything that is revealing at the TeV scale? Future and presumably not so distant future will show us what is correct.
 
\section{Is it compatible with Cosmology?}

The revolutionary development  of cosmology in recent years and the appearance of the Standard model of cosmology called the $\Lambda$CDM model~\cite{LCDM} allow one to compare the predictions of the Standard Model of particle physics with that of cosmology where they intersect. The main issues are:
\begin{itemize}
\item Baryon asymmetry of the Universe. The ratio of the number of baryons minus the number of anti-baryons in the Universe to the number of photons is given by an approximate formula~\cite{BAU}
\beq
\frac{N(B)-N(\bar B)}{N_\gamma}\sim (6.19\pm0.14)\times 10^{-10}.
\eeq
This number is still not explained in the SM and may require modification of the SM in future.
It requires larger CP-violation than in the strong sector of the SM giving some hints toward its lepton nature.
\item Relic abundance of the Dark Matter.  According to recent data from the Planck mission~\cite{Planck}, the energy balance of the Universe has the following shape
\beq Ordinary \ Matter =4.9 \%, \ Dark \ Matter=26.8 \%, \ Dark \ Energy=68.3 \%  \label{balance}\eeq

\noindent The problem of the Dark matter content is the problem of particle physics and seems to be beyond the SM. We will come to this point later. 
\item Number of neutrinos. The recent combined data from the Cosmic Microwave Background (CMB), the baryon acoustic oscillations (BAO), the WMAP polarization data (WP),  the Hubble Space Telescope (HST) and high-l temperature power spectrum (highL) give for the number of neutrinos the value~\cite{Planck}
\beq  N_{eff}(\nu)=3.52\pm 0.47\ \ at \ 95\% \ CL, \eeq
that well suits the SM with 3 generations assuming the quark-lepton symmetry.
\item Masses of neutrinos. From the same CMB, WP and HST   data plus  the gravitational lensing one gets the bound on the neutrino masses~\cite{MN}
\beq \sum m_\nu < 1.11(0.22) \ eV, \eeq
which is even stronger than in neutrino experiments. These extremely light neutrinos probably give us a hint
towards new physics responsible for their smallness like the see-saw mechanism and the Majorana nature of the neutrino.
\end{itemize}

\section{Hadron physics}
Looking back at the SM as the highest achievement in the description of matter we find some problems that were put aside in our race for the highest energy and intensity, namely, the problem of confinement and the problem of hot dense hadronic matter.

\subsection{Confinement and exotic hadrons}
\begin{figure}[ht]
\begin{center}
\leavevmode
\includegraphics[width=0.40\textwidth]{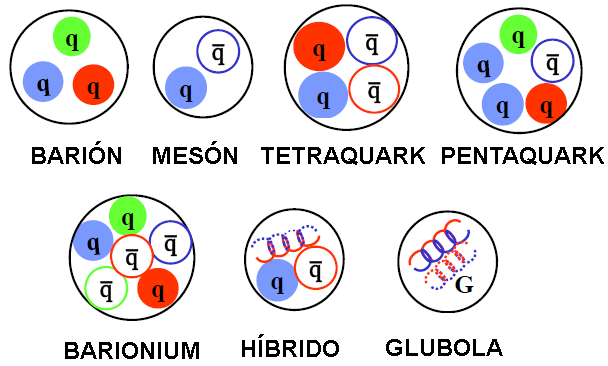}
\includegraphics[width=0.55\textwidth]{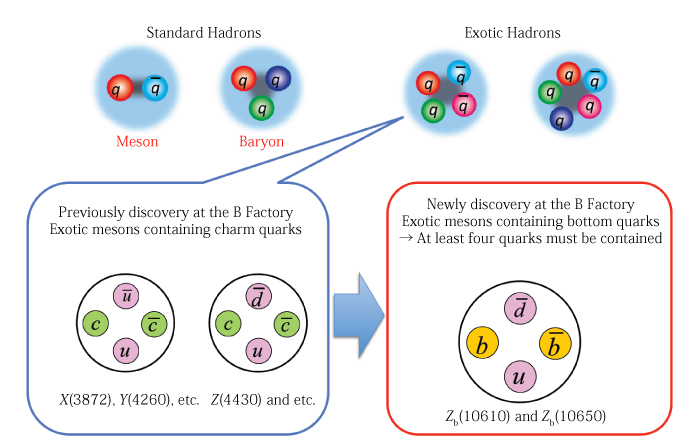}
\end{center}
\caption{Possible exotic colorless hadrons and newly discovered tetraquarks}
\label{exotic}
\end{figure}

The understanding of confinement is the challenging problem in particle physics well inside the SM.
Is it time to come back to it? We still do not understand how confinement actually works, why colorless states are the only observables, which bound states exist in Nature. Lattice calculations  seem to shed some light on it: we know that in mesons quark and anti-quark pairs are linked by the gluon string which 
has a tension and thus provides a linearly growing potential leading to confinement.  Trying to break this string one actually creates a new quark anti-quark pair thus again obtaining  colorless mesons. For baryons the situation is more sophisticated, the strings form the Mercedes-Benz star  with the same result as for mesons. However,  it is still unclear how these strings are formed and why they are restricted to the colorless states. And even if so, what about  other colorless states like the tetraquark, the pentaquark, the sextoquark, etc? Do they exist in Nature? (see Fig.\ref{exotic}~\cite{ExoticH}).
According to recent data, the pentaquark hadrons are at last unequivocally discovered at the LHC by the LHCb collaboration~\cite{PentaQ}. 

These new states require an adequate description probably within the  lattice gauge theories or within the
holographic approach or dual gauge theories. Or maybe we are back to analyticity and unitarity?

\subsection{Dense hadron matter}

\begin{figure}[ht]
\begin{center}
\leavevmode
\includegraphics[width=0.55\textwidth]{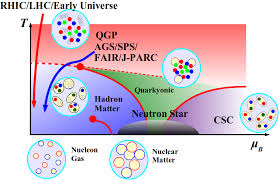}
\end{center}
\caption{The phase diagram of hadron matter}
\label{phase}
\end{figure}

Dense hadron matter might well be a new phase of matter with yet unknown properties which has no name so far.  It is known that at high density (high temperature) the usual description of hadron matter is not valid.
Hadrons do not exist above the Hagedorn temperature~\cite{HT}. What happens with a hadron gas at high pressure?
How to get the new phase? What is the relevant description? The popular phase diagram of hadron matter is shown in Fig.\ref{phase}~\cite{phase}. Here $T$ is the temperature and $\mu_B$ is the baryon chemical potential.
\begin{figure}[ht]
\begin{center}
\leavevmode
\includegraphics[width=0.55\textwidth]{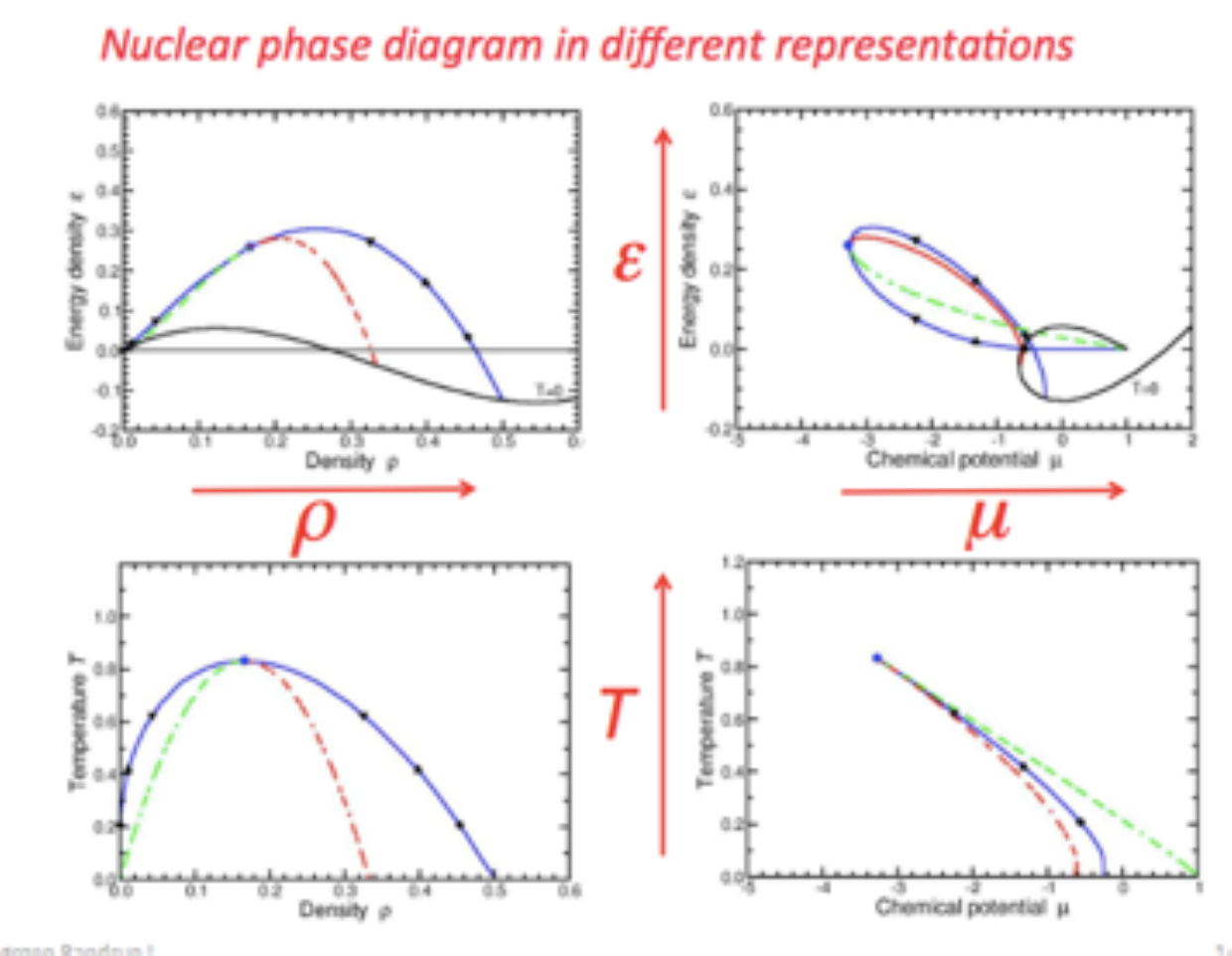}
\end{center}
\caption{The nuclear phase diagram in different representations }
\label{nuclear}
\end{figure} 

Usually, it is assumed that the phase diagram contains several phases with the phase transitions and critical points. The high temperature phase is usually referred to as a deconfinement one. To check whether it is true, one uses various methods including statistical mechanics,  nonequilibrium thermodynamics,
hydrodynamics,  and dual holographic models. There are several microscopic and macroscopic models~\cite{Brat}. As an example we show below (Fig.\ref{nuclear}) the nuclear phase diagram in different representations for different parameters~\cite{nuclphase}.
It represents rich new phenomena which  still have to be exploited.

\section{Search for new physics}
\subsection{The Higgs boson}
The Higgs boson still remains the target \#1 in search for new physics. And though there is no doubt that the  discovered particle is the CP-even scalar with all the properties of the Higgs boson, the main question remains: is it the SM Higgs boson or not? Are there alternatives to a single Higgs boson of the SM? The answer is positive. One may consider the singlet, doublet and triplet extensions of the SM, or their combinations~\cite{MS}.  The  guiding principle for these extensions is the custodial symmetry. It indicates that an approximate global symmetry exists,  broken by the vev to the diagonal ‘custodial’ symmetry  group $SU(2)_L\times SU(2)_R\to SU(2)_{L+R}$.  The custodial symmetry of the SM is responsible for the ratio
\beq
\rho=\frac{M_W^2}{M_Z^2\cos^2\theta_W}=1
\eeq
at the tree level. In the case of various extensions, when the Higgs field(s) transform under $SU(2)_L\times SU(2)_R$ as $\Phi\to L \Phi R$, the $\rho$-parameter can be constructed starting from the isospin and the hyper charge values of the Higgs multiplets~\cite{MS}
\beq
\rho=\frac{\sum\limits_{i=0}^{n}[I_i(I_i+1)-\frac 14 Y^2_i]v_i}{\sum\limits_{i=0}^{n}\frac 12 Y^2_iv_i}.
\eeq
For both $SU(2)$ singlet with $Y=0$ and $SU(2)$ doublet with $Y=\pm 1$ one has $\rho=1$. Moreover,  any number of singlets and doublets respect custodial symmetry at the tree level. This is not so for an arbitrary number of triplets.

\begin{figure}[ht]
\leavevmode  \hspace{0.7cm}\vspace{-0.4cm}

\includegraphics[width=0.40\textwidth]{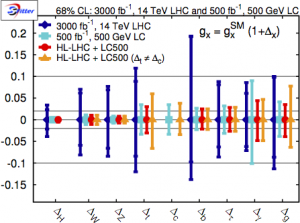}\vspace{-4.6cm}

\hspace{8.6cm}\includegraphics[width=0.41\textwidth]{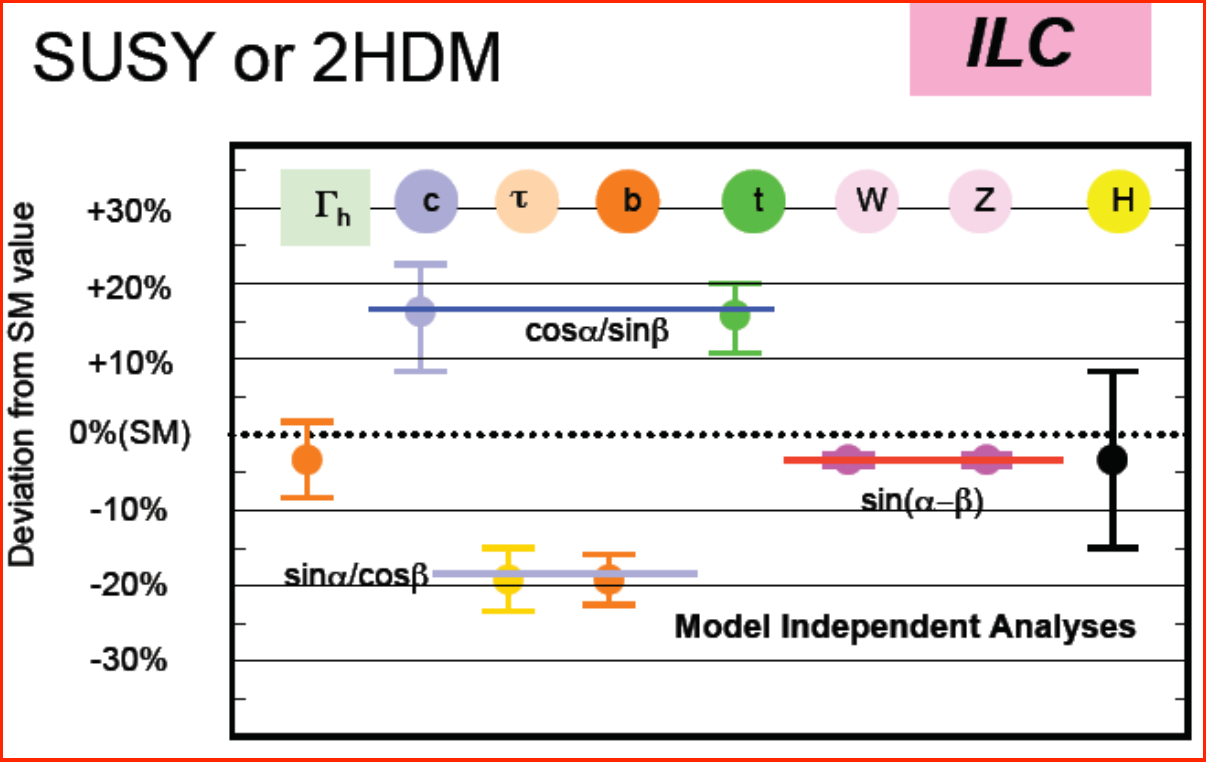}\vspace{0.4cm}
\caption{The accuracy of the measurement of the Higgs couplings  at various accelerators (left) and the required precision to distinguish SUSY from the 2HDM (right)}
\label{couplings}
\end{figure}

How  can one probe that  the Higgs boson is of the SM? There are two ways to do it. First of all, one has to probe the deviations from the SM Higgs couplings ( see Fig.\ref{couplings}~\cite{coup}). 

The name of the game is precision. At the few percent level one can distinguish, for example, the two Higgs doublet model of the MSSM type from the SM~\cite{coup2}.

\begin{figure}[ht]\vspace{0.6cm}
\leavevmode \hspace{1cm}
\includegraphics[width=0.30\textwidth, height=6cm]{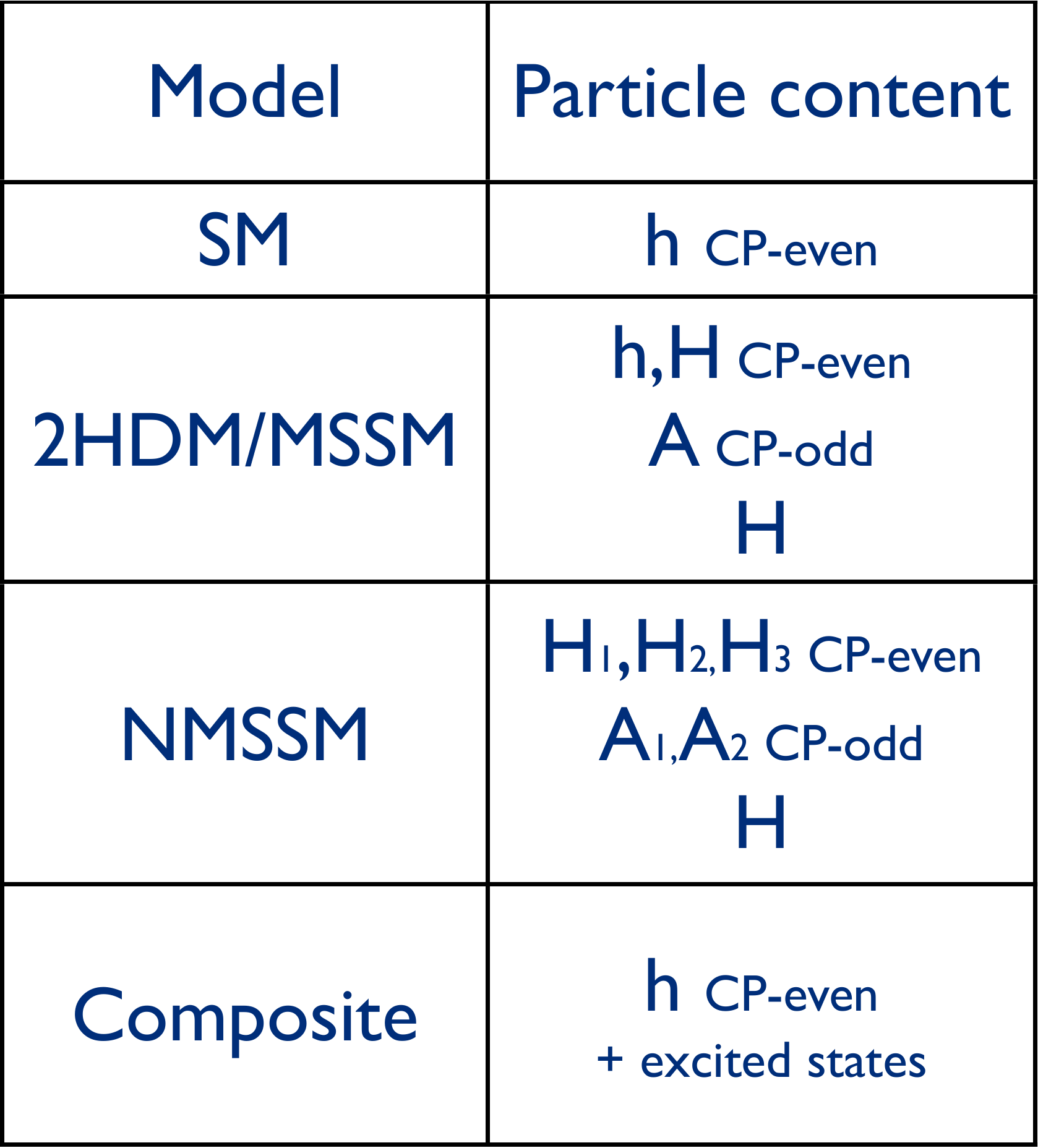} \vspace{-6.2cm}

\hspace{8.4cm}\includegraphics[width=0.34\textwidth,height=6.4cm]{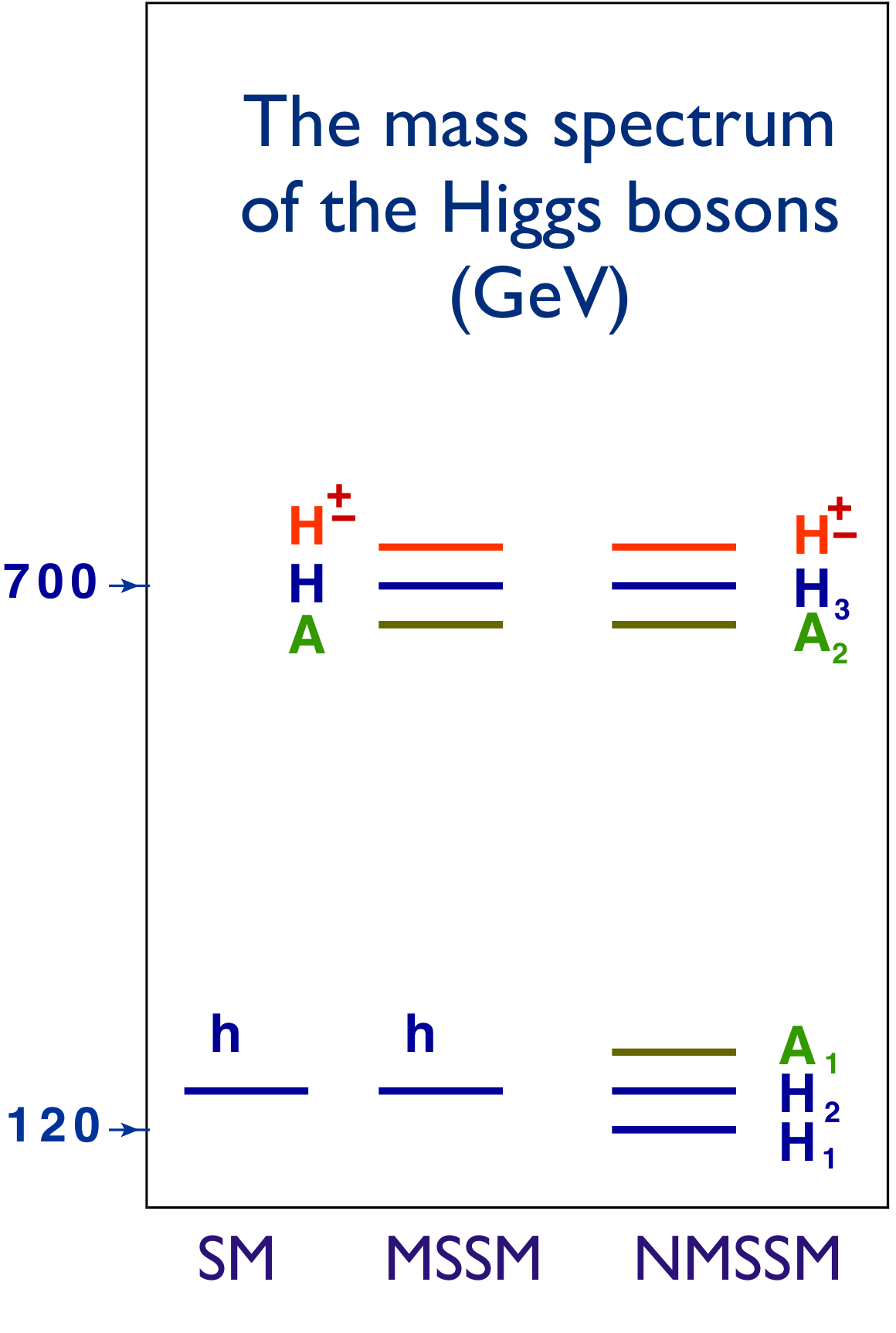}
\caption{The field content and the spectrum in various models of the Higgs sector}
\label{spectr}
\end{figure}

The second way is the direct search for additional scalars. In various extensions one can have extra CP-even, CP-odd and charged Higgs bosons. As an example, we present in Fig.\ref{spectr} the spectrum of the Higgs bosons in supersymmetric models (the MSSM  - two Higgs doublet model and the NMSSM - plus additional singlet).  
It may well be that we have found one of the light states that may not even be the lightest one.  The latter one may have very weak couplings and thus not being detectable~\cite{BBK}.

The Higgs physics has already started. This is the  task of vital importance to be fulfilled at the LHC but may  require an electron-positron collider.

\subsection{The Dark matter}	
Target \# 2 is the Dark matter. We know now that the amount of  the Dark matter in the Universe exceeds that of the usual matter by a factor of 5 (see eq.(\ref{balance}), but we do not know what it is made of. Some possible  candidates are:
\begin{itemize}
\item Macro objects  -- not seen in our Galaxy
\item New particles
\end{itemize}
\hspace*{2.8cm} -- heavy right neutrino \hspace{0.8cm} not favorable but possible\\
Not the SM $\left\{ 
\hspace*{0.1cm}\begin{minipage}{12cm}
\begin{tabular}{ll}
-- axion (axino) & might be invisible (?)\\
-- neutralino & detectable in three spheres\\
-- sneutrino & less theoretically favorable\\
-- gravitino  & might be undetectable (?) \\
 -- heavy photon  & possible \\
-- heavy pseudo-goldstone & but not related\\
-- light sterile Higgs & to  other models
\end{tabular}
\end{minipage} \right.$\vspace{0.3cm}

\begin{figure}[ht]
\leavevmode
\includegraphics[width=0.53\textwidth]{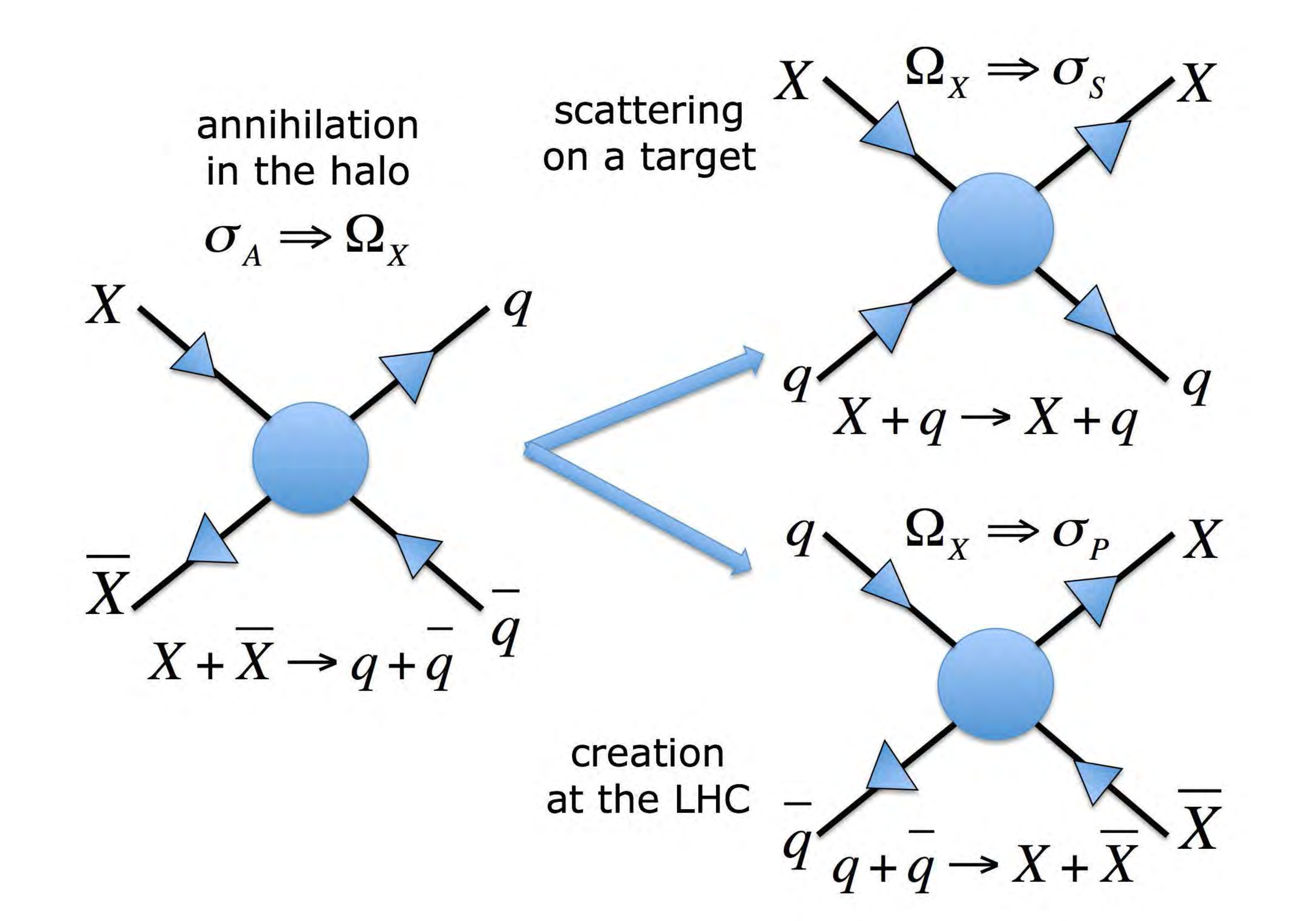}\hspace{1.2cm}
\includegraphics[width=0.30\textwidth]{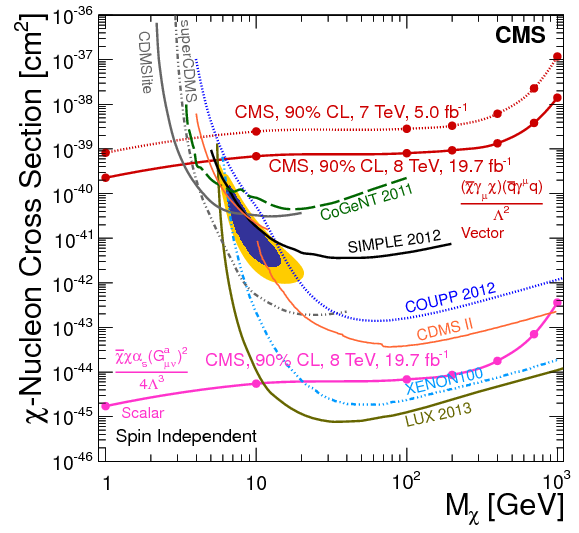}
\caption{Detection of WIMPs in there different spheres (left) and experimental data for the underground experiments and accelerators (right)}
\label{darkdetection}
\end{figure}

Our best chance to detect the Dark matter particle would be via the weak interaction. The so-called WIMP (weakly interacting massive particle) can be simultaneously detected in three spheres: via annihilation in the halo of our Galaxy (irregularities in cosmic ray spectra), via scattering on a target in underground experiments (recoil of a nuclei) and via creation at accelerators (missing energy) (see Fig.\ref{darkdetection} (left)~\cite{RKolb}).
This search is already on the way with a negative result so far. The typical plot is shown in Fig.\ref{darkdetection}(right)~\cite{DarkM}
where the results of the direct search and the collider experiments are presented. 
One can see the complimentary nature of these studies with the advantage of the accelerators at low masses and the advantage of the underground experiments at higher masses of WIMPs. The latter has already almost reached the neutrino floor where the background of neutrinos will be prevailing~\cite{floor}. Apparently, WIMPs are our chance though we have to look elsewhere.

\subsection{Supersymmetry}
Supersymmetry is an obvious target \#3.  Supersymmetry is a dream of a unified theory of all particles and interactions~\cite{SUSY}.
\begin{figure}[ht]
\begin{center}
\leavevmode
\includegraphics[width=0.95\textwidth]{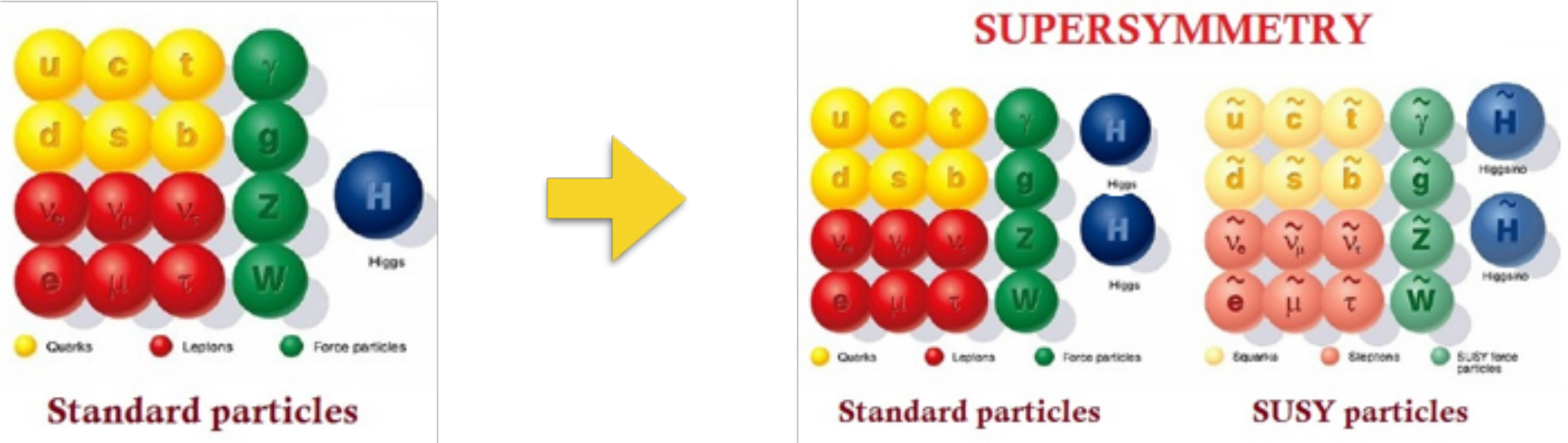}
\end{center}
\caption{Particle content of Minimal SUSY model}
\label{SUSY}
\end{figure} 
Supersymmetry remains, to this date, a well-motivated, much anticipated extension to the Standard Model of particle physics~\cite{BT}. 

\begin{figure}[ht]
\begin{center}
\leavevmode
\includegraphics[width=0.13\textwidth]{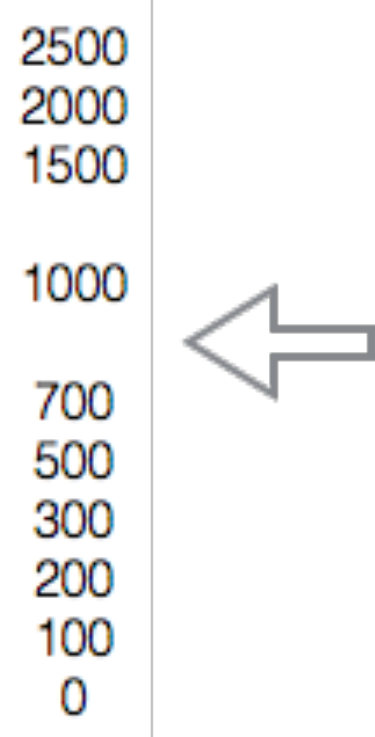}
\includegraphics[width=0.38\textwidth]{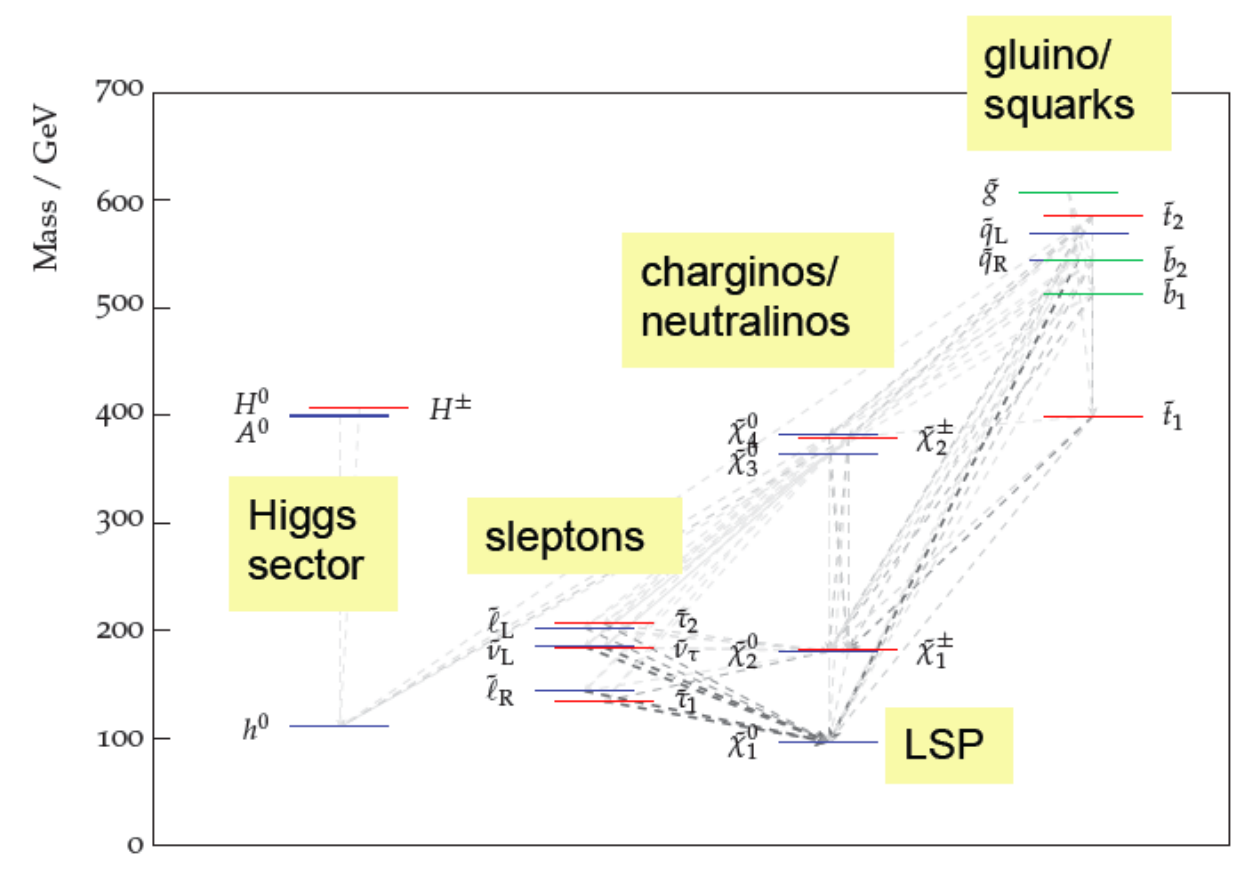}
\end{center}
\caption{The typical SUSY mass spectrum} 
\label{SUSYspec}
\end{figure}

With the advent of the LHC a huge new ground of SUSY masses is within reach. However,  a search is defined by its signature and by its background  estimation method. Still, if SUSY is the answer to the “naturalness” problem, then there must exist light colored particles. The typical spectrum of SUSY particles consistent with the naturalness paradigm in shown in Fig.\ref{SUSYspec}~\cite{spectr}. At the left, it is shown how the scale of SUSY searches has shifted after the first run of the LHC.
	
	Many available supersymmetric models differ mostly by the way of supersymmetry breaking. 
Since this problem has not found its obvious solution, one is left with the phenomenological set of parameters motivated by either the simplification of parameter space, like in the MSSM with universality requirement,  or the restricted  number of experimental signatures, like in the so-called simplified models.

\begin{figure}[htb]
\begin{center}
\leavevmode
\includegraphics[width=0.40\textwidth]{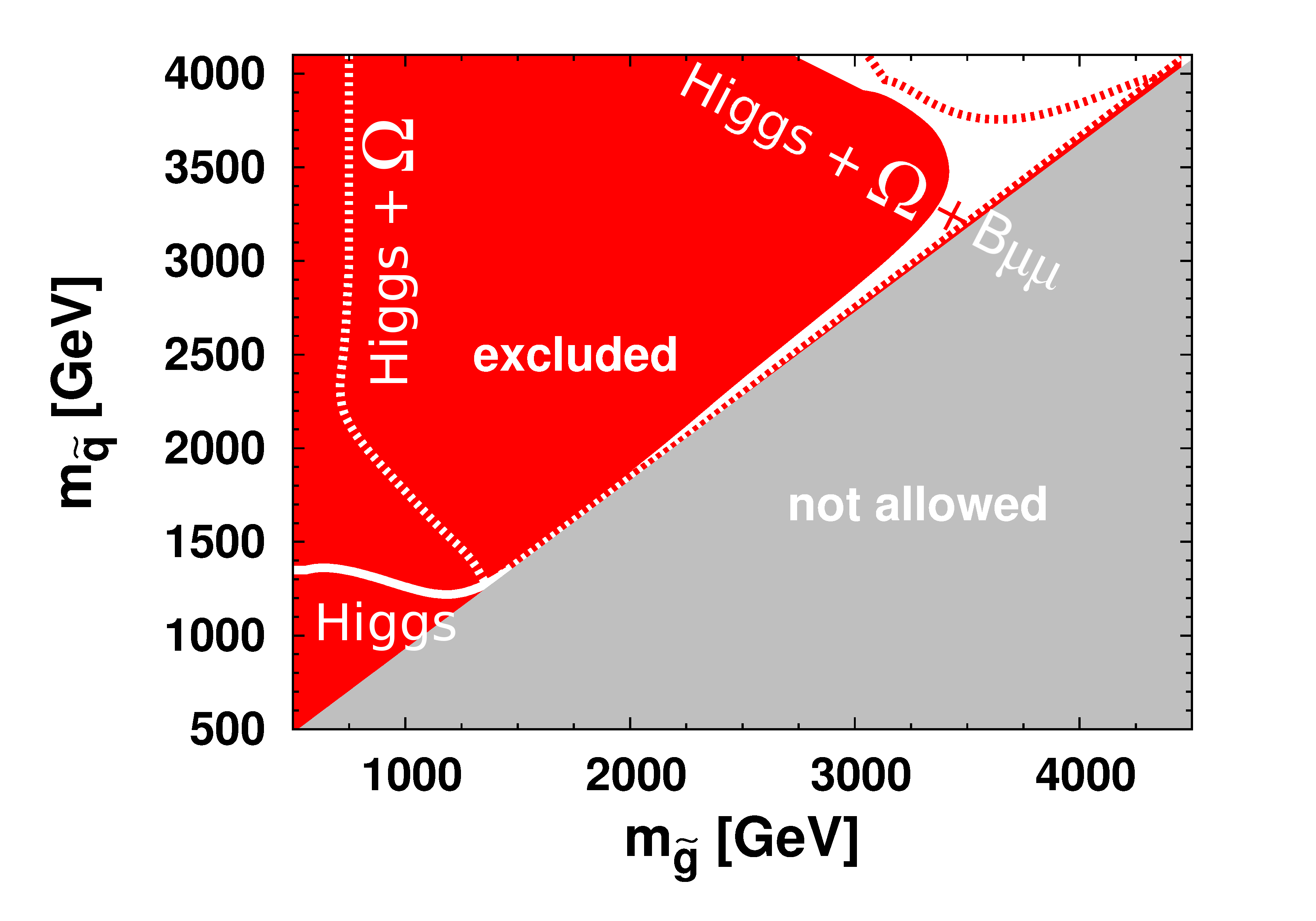}
\includegraphics[width=0.40\textwidth]{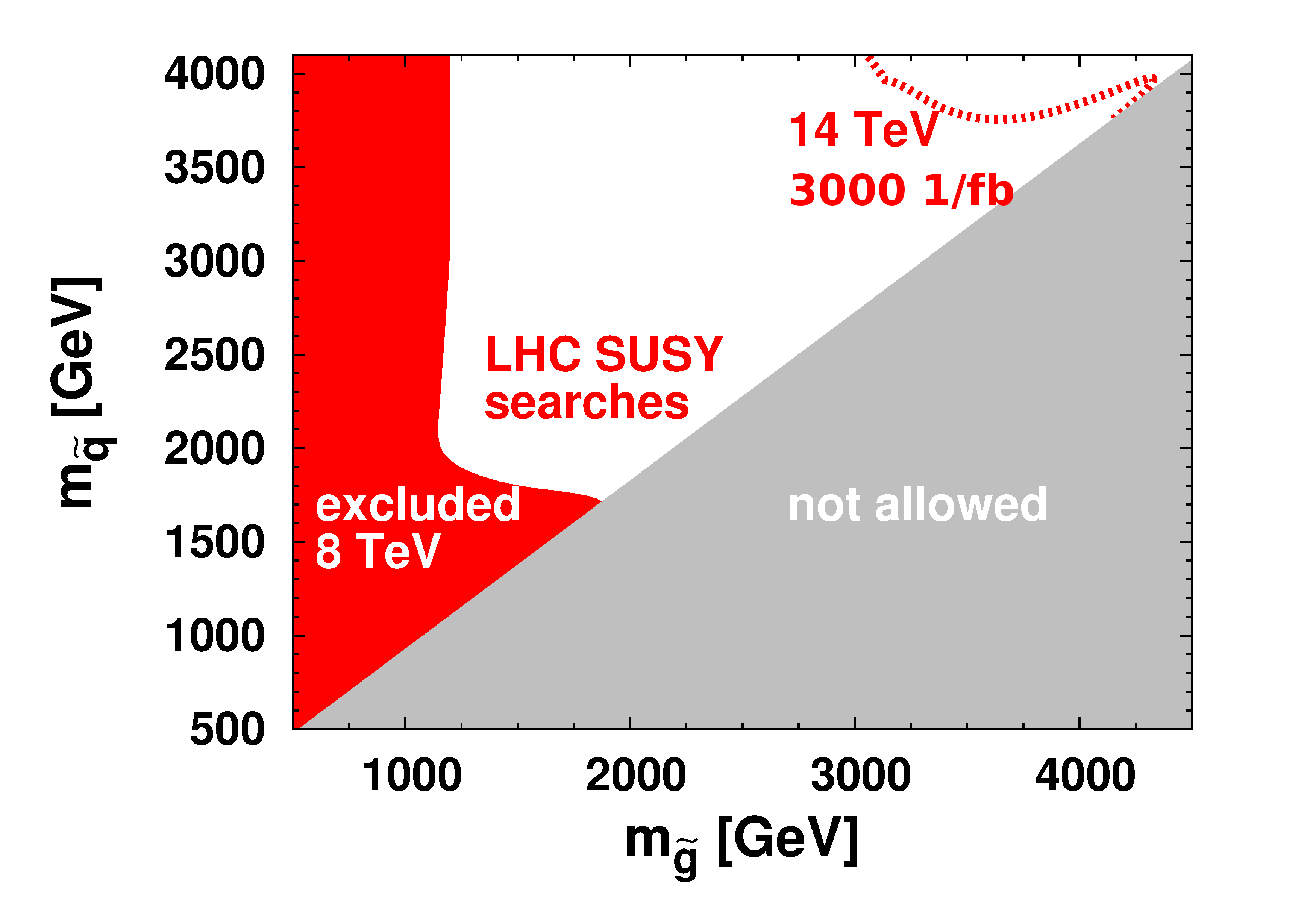}

CMSSM \hspace{6cm} NMSSM
\end{center}
	\caption{The SYSY reach of the LHC  in the SUSY mass plane for the CMSSM (left) and NMSSM (right)} 
\label{limits}
\end{figure}

In both the cases the experimental data on direct SUSY searches and the indirect SUSY contributions to rare decays, relic dark matter abundance, the lightest Higgs mass, etc push the limits on SUSY masses to a few TeV scale~\cite{BBK2}), which makes the observation more problematic. Moreover, pushing the SUSY threshold even further, we start losing the main motivation for a low energy SUSY, namely the solution of the hierarchy problem and unification of the gauge couplings. Note, however, the conclusions crucially depend on the model applied, as one may see from Fig.\ref{limits} below~\cite{BBK}.  Going from  the MSSM to the NMSSM, for example, allows one to incorporate the 125 GeV Higgs mass and still keep the light super partners. 

 The absence of a model independent way of predictions and analysis makes it difficult to put strict limits on the low energy supersymmetry. However, there is a crucial moment now: either we find SUSY at the LHC eventually or we might have no other chance. Then we have to solve the hierarchy problem some other way! (which way?). 
 
\subsection{Extra Dimensions/ Exotics}

\begin{figure}[ht]
\begin{center}
\leavevmode
\includegraphics[width=0.50\textwidth]{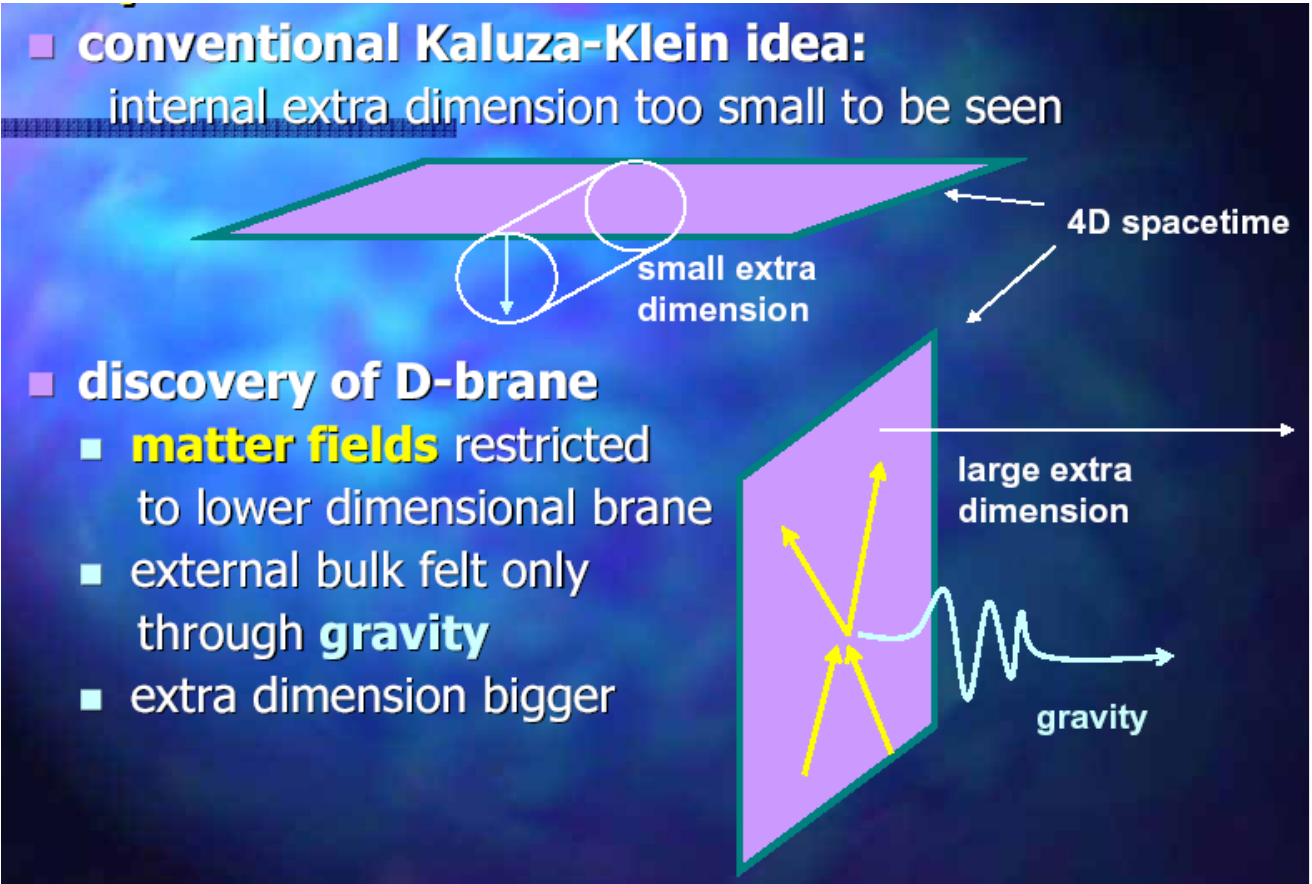}
\end{center}
\caption{Compact or large extra dimensions scenarios} 
\label{extraD}
\end{figure}

The extra dimensional approach might be an alternative to low energy supersymmetry or  might also include SUSY within the brane world framework. Usually, the two main versions of extra dimensions are considered: the compact extra dimensions a la Kaluza-Klein picture  (the ADD scenario~\cite{ADD}) or the large extra dimensions (the Randall-Sandrum scenarios~\cite{RS}).
Schematically, they are shown in Fig.\ref{extraD}~\cite{ExtraD}.

These kinds of models demonstrate a significant departure from the Standard Model since they not only contain the new fields and interactions but the whole framework of renormalizable quantum field theory is left behind. Apparently,  this approach requires a new technique which is still to be developed. I would present my view on the extra dimensional brane world scenario in the form of a dialogue.

Q:  Do we really live on a brane? 

A: We have to check it.

Q: Do we have good reasons to believe in it?

A: No, but it is appealing.

Q: Why $D>4$?

A: String theory loves it.

Q: Is it what we believe in?

A: We believe in BIG deal!


The phenomenology of extra dimensions is quite rich, though it is not linked to any particular scale. Possible experimental manifestations  include:
search for $Z'$ (Di-muon events),
search for $W'$ (single muon/ jets),
search for a resonance decaying into t-tbar,
search for diboson resonances,
search for monojets + invisible particles.

Besides extra dimension there are a lot of exotic possibilities. None of them have been found so far, though 
one cannot a priori say what is realized in Nature. Some common topics are listed below:
Leptoquarks,  long-lived particles,  off-pointing photons,  excited fermions, 
contact interactions, etc.  The drawback of all these approaches is the lack of real motivation and hence the arbitrariness of the scale of new physics.

\subsection{Compositeness}

\begin{figure}[ht]
\leavevmode
\hspace{4cm}\includegraphics[width=0.25\textwidth]{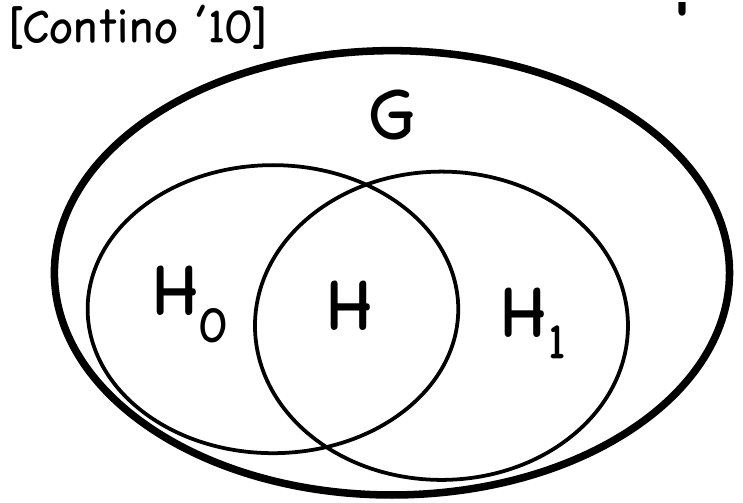}\vspace{-2cm}

\hspace{10cm}Global symmetry G 

\hspace{10cm}broken to H of the SM\vspace{1.0cm}
\caption{Breaking of the global group G down to the SM subgroup H} 
\label{group}
\end{figure}

Compositeness is in a sense a natural continuation of the chain of particle physics starting from an atom and going down to quarks. The question is: moving to higher energies or smaller distances do we have to stay with the same fundamental particles or the new level appears.  Answering this question we first of all look at the Higgs boson  as an obvious analogy with the $\pi$-meson as a pseudo Nambu-Goldstone boson of the chiral symmetry. One has in mind the construction when some global symmetry group $G$ is broken down to the symmetry group of the Standard Model H (see Fig.\ref{group})~\cite{composit}.

As a result, the Higgs boson becomes the pseudo Nambu-Goldstone particle like the $\pi$-meson, and the $W$ and $Z$ bosons have an analogy with the vector $\rho$-meson. There should also be exited states
like $\pi', \pi'',\rho',\rho''$, etc.

The advantage of this approach is that there is  no artificial scalar field, everything is dictated by the symmetry group.  The masses of these states are protected from high energy physics contribution, thus eliminating the  hierarchy problem.

One can go even further and consider quarks and leptons also as  composite states made of some {\it preons}~\cite{preons}. This would require new strong confining interactions. In earlier days, these types of models were referred to as technicolor, or walking technicolor, or extended technicolor. They have their own problems
and got new development now~\cite{TCnow}.  The drawback of these models is the absence of excited states so far, the problems with the EW phenomenology and the absence of a  viable  simple  scheme. Still this approach has the right to exist.

\section{Concluding remarks}
The LHC experiments are at the front line of  a mystery land. We make the first attempts to look beyond the horizon. We have to be persistent and have to be  patient.  The main goals are:
\begin{itemize}
\item[$\blacktriangleright$]  Target \#1: The Higgs sector;
\item[$\blacktriangleright$]  Target \#2: The Dark Matter;
\item[$\blacktriangleright$]  Target \#3: The New physics (supersymmetry);
\end{itemize}
In attempts to achieve these goals one should have in mind that
\begin{itemize}
\item The future development of HEP crucially depends on the LHC outcome;

\item Complimentary searches for dark matter and insights in neutrino physics are of extreme importance;

\item The areas that were left behind come to the front: confinement, exotic hadrons, dense hadron matter.
\end{itemize}
I bet that discoveries will come!
\vspace{0.3cm}

{\bf Acknowledgments}
I am grateful to the Organizing Committee of the LHCp Conference for a challenge to give this talk and  for a warm hospitality at St.Petersburg.



\begin{thebibliography}{50}

\bibitem{Landau}  L.D. Landau, On quantum field theory,  in "Niels Bohr and the Development of Physics",  London: Pergamon Press, 1955;
\bibitem{Anomaly}  S. Adler, Axial-Vector Vertex in Spinor Electrodynamics, {\em Phys. Rev.} {\bf 177} (1969) 2426 ; 
J.S. Bell and R. Jackiw, A PCAC puzzle: $\pi^0\to \gamma\gamma$ in the $\sigma$-model, {\em Nuovo Cimento} {\bf 60A} (1969)  47;
S. Adler and W.A. Bardeen,  Corrections in the Anomalous Axial-Vector Divergence Equation, {\em Phys. Rev.} {\bf 182} (1969) 157.

\bibitem{AnomSM}  M.Peskin and D.Schreder, "An Introduction to Quantum Field Theory", Addison-Wesley Pub. Company, 1995.

\bibitem{vacuum}  G. Degrassi et al,  Higgs mass and vacuum stability in the Standard Model at NNLO, {\em JHEP} {\bf 1208}  (2012) 098, e-Print: arXiv:1205.6497.

\bibitem{highorder}	J. R. Espinosa, Vacuum Stability and the Higgs Boson,  
PoS LATTICE2013 (2014) 010,  e-Print: arXiv:1311.1970.

\bibitem{WB}  J. Wess, J. Bagger, "Supersymmetry and Supergravity", Princeton Univ. Press, 1983.

\bibitem{Kazakov}  A.V. Gladyshev, D.I. Kazakov, Is (Low Energy) SUSY Still  Alive?,  Lectures at ESHEP-2012, 
CERN-2014-008.107,  e-Print: arXiv:1212.2548.

\bibitem{Littlehierarchy} A. Birkedal, Z. Chacko, M. K. Gaillard,  Little Supersymmetry and the Supersymmetric Little Hierarchy Problem, {\em JHEP} {\bf 0410} (2004) 036, e-Print: arXiv:hep-ph/0404197.

\bibitem{RS}  L. Randall and R. Sundrum, Large Mass Hierarchy from a Small Extra Dimension, {\em Phys. Rev. Lett.} {\bf 83} (1999) 3370, e-print: hep-ph/9905221. L. Randall and R. Sundrum, An Alternative to Compactification, {\em Phys. Rev. Lett.} {\bf 83} (1999) 4690, e-print: hep-th/9906064.

\bibitem{EWPool}  G-fitter, A Generic Fitter Project for HEP Model Testing, http://project-gfitter.web.cern.ch; PDG, {\em Chin.Phys.} {\bf 38} (2014) 090001.
\bibitem{Flavor} CKM-fitter,   http://ckmfitter.in2p3.fr, PDG, {\em Chin.Phys.} {\bf 38} (2014) 090001.
\bibitem{PTCalc}  Talks at the LHCp2015 Conference: K. Melnikov,  Calculating Higgs boson properties in SM;
A. Vicini, General status and future prospects of HO corrections; S.Forte,
Perturbative QCD at the LHC; S.A.Pozzorini,
Electroweak theory at the LHC.
\bibitem{g2}  A. Hoecker  and W.J. Marciano,  The muon anomalous magnetic moment,  http://pdg.lbl.gov/2013/reviews/rpp2013-rev-g-2-muon-anom-mag-moment.pdf
\bibitem{Vub}  LHCb Collaboration, Determination of the quark coupling strength $ |Vub|$ using baryonic decays, {\em Nature Physics} {\bf 11} (2015) 743, http://www.nature.com/nphys/journal/v11/n9/full/nphys3415.html 
\bibitem{axion}  J. E. Kim, Constraints on very light axions from cavity experiments,  {\em Phys. Rev.} {\bf D58} (1998) 055006.
\bibitem{rare}  Talks at the LHCp2015 Conference: C. Bobeth (for CMS and LHCb Collaborations), Theoretical perspective on rare and semi-rare B decays.
\bibitem{PartonD}  P. Mulders and R. Tangerman, The complete tree-level result up to order 1/Q for polarized deep-inelastic leptoproduction, {\em Nucl.Phys.} {\bf B461}  (1996) 197, e-print: hep-ph/9510301.
R. Angeles-Martinez et al, 
Transverse momentum dependent (TMD) parton distribution functions: status and prospects, e-Print: arXiv:1507.05267
\bibitem{GIM}   S.L. Glashow, J. Iliopoulos, L. Maiani,  Weak Interactions with LeptonРHadron Symmetry, {\em Phys. Rev.} {\bf D2 (7)} (1970) 1285. 
\bibitem{scale} C. Quigg, The Coming Revolutions in Particle Physics, {\em Scientific American}, February, 2008;
 http://chemphys.armstrong.edu
\bibitem{LCDM}   A. Linde,  "An Introduction to Modern Cosmology" (2nd ed.). London: Wiley, 2003;
S. Weinberg,  "Cosmology", Oxford University Press, Oxford, (2008).
\bibitem{BAU}  E. W. Kolb and M. S. Turner, "The Early Universe", Addison-Wesley (1990).
\bibitem{Planck} Planck 2013 results. XVI. Cosmological parameters, {\em Astronomy and Astrophysics} {\bf 571} (2014) A16, e-print: arXiv:1303.5076.
\bibitem{MN}  Jian-Wei Hu, Rong-Gen Cai, Zong-Kuan Guo, Bin Hu, Cosmological parameter estimation from CMB and X-ray clusters after Planck, {\em JCAP} {\bf 1405} (2014) 020,  e-print: arXiv:1401.0717.
\bibitem{ExoticH} F.R.Villatoro, http://francis.naukas.com/2011/10/08/que-paso-con-los-pentaquarks
http://phys.org/news/2012-01-belle-heavy-exotic-hadrons.html
\bibitem{PentaQ}  R. Aaij et al. (LHCb Collaboration), Observation of $J/\psi p$ resonances consistent with pentaquark states in $\Lambda^0_b\to J/\psi K^-p$ decays, {\em Phys. Rev. Lett.} {\bf 115} (2015)  072001.
\bibitem{HT} J.Cleymans and D. Worku, The Hagedorn temperature Revisited, {\em Mod. Phys. Lett.} {\bf A26} (2011) 1197; e-print: arXiv: 1103.1463.
\bibitem{phase}  A. Ohnishi, Phase diagram and heavy-ion collisions: Overview,
{\em Prog.Theor.Phys.Suppl.} {\bf 193} (2012) 1, e-Print: arXiv:1112.3210 [nucl-th].
\bibitem{Brat} E. Bratkovskaya,  Microscopic dynamical models for heavy ion collisions, Lectures at International Summer School "Dense Matter 2015", JINR, Dubna, July 2015, http://theor.jinr.ru/~diastp/dm15/
\bibitem{nuclphase}  J. Randrup, Spinodal instabilites at the deconfinement phase transition,  Lectures at  International Summer School "Dense Matter 2015", JINR, Dubna, July 2015,  http://theor.jinr.ru/~diastp/dm15/
\bibitem{MS} M.Spannowsky, Higgs Phenomenology, Lectures at Helmholtz - DIAS International Summer School 
"Theory challenges for LHC physics,"  JINR, Dubna,  July 2015, http://theor.jinr.ru/~calc2015/
\bibitem{coup} S-Fitter Collaboration, \\
http://groups.lal.in2p3.fr/atlas/files/2013/01/HiggsComparisonLC500.png
\bibitem{coup2}S. Yamashita,  Physics at International Linear Collider,  7th ACFA WS, http://hep1.phys.ntu.edu.tw/.../P2-2-Yamashita.pdf 
\bibitem{BBK}  C. Beskidt, W. de Boer, D.I. Kazakov,  A comparison of the Higgs sectors of the CMSSM and NMSSM for a 126 GeV Higgs boson,  {\em Phys. Lett.} {\bf B726} (2013) 758, \  e-Print: arXiv:1308.1333 [hep-ph] 
\bibitem{RKolb} E.W.Kolb,  A Dark Universe: Dark Matter and Dark Energy , CERN Academic Lectures, 
www.infocobuild.com/.../cosmology-cern.htm
\bibitem{DarkM} D. d'Enterria, (CMS Collaboration), CMS physics highlights in the LHC Run 1, PoS Bormio2015 (2015) 027, e-Print: arXiv:1504.06519.
\bibitem{floor} P.Grothaus, M. Fairbairn, J. Monroe, Directional Dark Matter Detection Beyond the Neutrino Bound
{\em Phys. Rev.} {\bf D 90} (2014)  055018, e-print: arXiv:1406.5047 [hep-ph].
\bibitem{SUSY} P. Fayet and S. Ferrara, Supersymmetry, {\em Phys. Rep.} {\bf 32} (1977) 249;\\
M. F. Sohnius, Introducing Supersymmetry, {\em Phys. Rep.} {\bf 128} (1985) 41;\\
H. P. Nilles, Supersymmetry, supergravity and particle physics, {\em Phys. Rep.} {\bf 110} (1984) 1; 
H. E. Haber and G. L. Kane, The search for supersymmetry: Probing physics beyond the standard model, {\em Phys. Rep.} {\bf 117} (1985) 75;\\
A. B. Lahanas and D. V. Nanopoulos, The road to no-scale supergravity, {\em Phys. Rep.} {\bf 145} (1987) 1.
\bibitem{BT} H. Baer, X. Tata, "Weak Scale Supersymmetry", Cambridge University Press, 2006.
\bibitem{spectr} M.Monaco, M. Pierini, A. Romanino and M. Spinrath, Phenomenology of Minimal Unified Tree Level Gauge Mediation at the LHC, {\em JHEP} {\bf 1307} (2013) 078, arXiv:1302.1305 [hep-ph].
\bibitem{BBK2} C. Beskidt, W. de Boer, D.I. Kazakov, Where is SUSY?  {\em JHEP} {\bf 1205} (2012) 094, 
e-Print: arXiv:1202.3366. 
\bibitem{ADD} N. Arkani-Hamed, S. Dimopoulos and G. Dvali, {\em Phys. Lett.} {\bf B429} (1998) 263, e-print: hep- ph/9803315;
N. Arkani-Hamed, S. Dimopoulos and G. Dvali, {\em Phys.Rev.} {\bf D59} (1999) 086004, e-print: hep- ph/9807344.
\bibitem{ExtraD} R.Maartens, Brane Cosmology, Rencontres de Moriond, Electro Weak Interactions and Unified Theories 2004, http:/moriond.in2p3.fr/J04/trans/maartens.pdf
\bibitem{composit} R.Contino, The Higgs as a Composite Nambu-Goldstone Boson, Theoretical Advanced Study Institute in Elementary Particle Physics : Physics of the Large and the Small. (TASI 2009), e-Print: arXiv:1005.4269.
\bibitem{preons} I.A. D'Souza, C.S. Kalman,  "Preons: Models of Leptons, Quarks and Gauge Bosons as Composite Objects", World Scientific, 1992
\bibitem{TCnow} A. Belyaev, M.S. Brown, R. Foadi, M.T. Frandsen, The Technicolor Higgs in the Light of LHC Data 
{\em Phys.Rev.} {\bf D90} (2014) 035012 , e-Print: arXiv:1309.2097; 
M. Antola, S. Di Chiara, K. Tuominen, Ultraviolet Complete Technicolor and Higgs Physics at LHC,
{\em Nucl.Phys.} {\bf B899} (2015) 55,  e-Print: arXiv:1307.4755. 
\end{thebibliography}
\end{document}